\documentclass[aps,prl,twocolumn,superscriptaddress]{revtex4-2}
\usepackage{graphicx}
\usepackage{amssymb}
\usepackage{amsmath}
\usepackage{color}
\draft

\begin{document}

\preprint{ESR on k-AgCN}

\title{Gapped magnetic ground state in the spin-liquid candidate $\kappa$-(BEDT-TTF)$_2$Ag$_2$(CN)$_3$\\ suggested by
 magnetic spectroscopy }

\author{Sudip Pal}
\affiliation{1. Physikalisches Institut, Universität Stuttgart, Pfaffenwaldring 57, 70569 Stuttgart Germany}
\author{Bj{\"o}rn Miksch}
\affiliation{1. Physikalisches Institut, Universität Stuttgart, Pfaffenwaldring 57, 70569 Stuttgart Germany}
\author{Hans-Albrecht Krug von Nidda}
\affiliation{Experimental Physics V, Center for Electronic Correlations and Magnetism, Universit\"at Augsburg, Universitätsstraße 1, 86159 Augsburg, Germany}
\author{Anastasia Bauernfeind}
\affiliation{1. Physikalisches Institut, Universität Stuttgart, Pfaffenwaldring 57, 70569 Stuttgart Germany}
\author{Marc Scheffler}
\affiliation{1. Physikalisches Institut, Universität Stuttgart, Pfaffenwaldring 57, 70569 Stuttgart Germany}
\author{Yukihoro Yoshida}
\affiliation{Faculty of Agriculture, Meijo University, Nagoya, 468-8502, Japan}
\affiliation{Division of Chemistry, Graduate School of Science, Kyoto University, Kyoto, 606-8502, Japan}
\author{Gunzi Saito}
\affiliation{Faculty of Agriculture, Meijo University, Nagoya, 468-8502, Japan}
\affiliation{Division of Chemistry, Graduate School of Science, Kyoto University, Kyoto, 606-8502, Japan}
\author{Atsushi Kawamoto}
\affiliation{Department of Science, Hokkaido University, Sapporo, 060-0810, Japan}
\author{C{\'e}cile M{\'e}zi{\`e}re}
\affiliation{Univ Angers, CNRS, MOLTECH-Anjou, SFR MATRIX, 49000 Angers, France}
\author{Narcis Avarvari}
\affiliation{Univ Angers, CNRS, MOLTECH-Anjou, SFR MATRIX, 49000 Angers, France}
\author{John A. Schlueter}
\affiliation{Material Science Division, Argonne National Laboratory, Argonne, IL 60439-4831, U.S.A.}
\affiliation{National Science Foundation, Alexandria, VA 22314, U.S.A.}
\author{Andrej Pustogow}
\affiliation{Institute of Solid State Physics, TU Wien, 1040 Vienna, Austria}
\author{Martin Dressel}
\affiliation{1. Physikalisches Institut, Universität Stuttgart, Pfaffenwaldring 57, 70569 Stuttgart Germany}


\begin{abstract}
The nature of the magnetic ground state of highly frustrated systems remained puzzling to this day. Here, we have performed multifrequency electron spin resonance (ESR) measurements on a putative quantum spin liquid compound $\kappa$-(BEDT-TTF)$_2$Ag$_2$(CN)$_3$,
which is a rare example of $S = 1/2$ spins on a triangular lattice.
At high temperatures, the spin susceptibility exhibits a weak temperature dependence which can be described by the Heisenberg model with an antiferromagnetic exchange interaction of strength $J/k_B \approx 175$~K. At low temperatures, however, the rapid drop of the static spin susceptibility, together with monotonic decrease of the ESR linewidth indicates that strong singlet correlations develop below a pairing energy scale $T^*$  accompanied by a spin gap. On the other hand, a weak Curie-like spin susceptibility and the angular dependence of the linewidth suggest additional contribution from impurity spins.
We propose the gradual formation of spin singlets with an inhomogeneous spin gap at low temperatures.
\end{abstract}
\date{\today}
\maketitle

The so-called quantum spin liquid state (QSL) is an illusive magnetic phase \cite{Zhou2017,Broholm2020} and its fundamental properties are extensively discussed in condensed matter physics at present. Initiated by the proposal of a resonating valence-bond state and its subsequent connection to high-temperature superconductivity \cite{Anderson1973,*Fazekas1974,Anderson1987}, different classes of materials have been scrutinized in search of QSL states, starting from the organic Mott insulators to various inorganic materials such as Yb$^{3+}$ based compounds \cite{Scheie2023}, Ba$_3$CoSb$_2$O$_9$ \cite{Shirata2012,Susuki2013}, Cu-based kagome-lattice systems such as Herbertsmite and related compounds \cite{Khuntia2020}.
In this context, the ground state of the organic compounds $\kappa$-(BEDT-TTF)$_2$Cu$_2$(CN)$_3$ (abbreviated as $\kappa$-CuCN in the following) and $\beta^{\prime}$-EtMe$_3$Sb[Pd(dmit)$_2$]$_2$ were intensely studied over the last decade \cite{Shimizu2003,Itou2010,Shimizu2016,Dressel2020}. In case of $\kappa$-CuCN, below a characteristic temperature $T^* \approx 6$~K, significant lattice and sound-velocity anomalies have been observed \cite{Manna2010,Poirier2014}. The spin susceptibility exhibits a rapid decay, which unambiguously indicates the opening of an energy gap in the spin excitation spectrum below $T^*$ \cite{Padmalekha2015,Miksch2021,Pustogow2022}. The low-entropy nature of the ground state has been further corroborated by the negative slope of the insulator-metal boundary in the $T$-$p$ phase diagram and negative magneto-resistance around the Mott transition \cite{Pustogow2018,Pustogow2023,Kawasugi2023}. Despite these
compelling experimental evidences that $\kappa$-CuCN forms a valence-bond solid ground state below $T^*$, the debate has not completely deceased \cite{Riedl2019,Matsuura2022,Riedl2022,Liebman2024}.

\begin{figure}[b]
\centering
\includegraphics[width=\columnwidth]{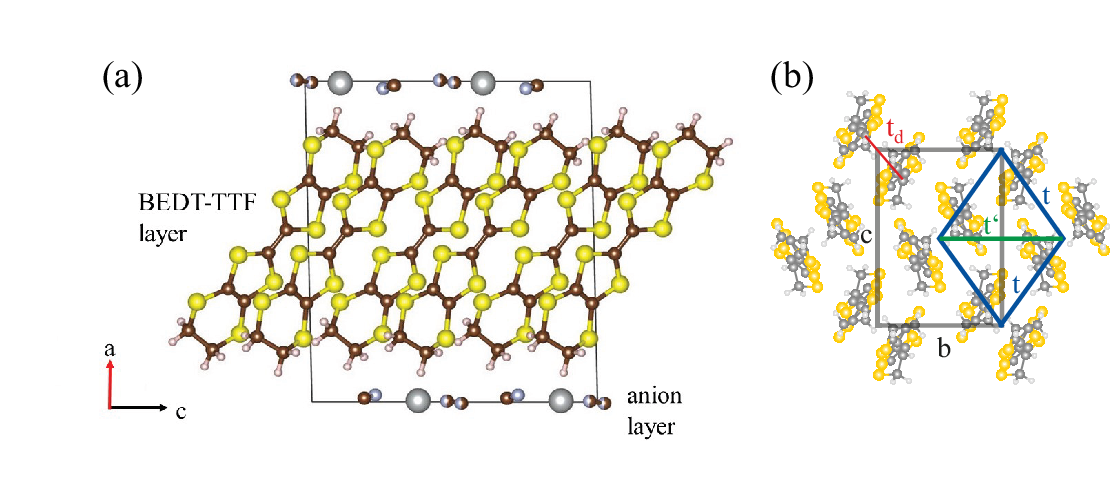}
\caption{Crystal structure of $\kappa$-(BEDT-TTF)$_2$\-Ag$_2$(CN)$_3$.
(a)~The organic (BEDT-TTF) layers and inorganic anion sheets alternate along the $a$-axis with the BEDT-TTF molecules tilted in $c$-direction. (b)~In the $bc$-pane, the (BEDT-TTF)$_2$ dimers form an almost perfect triangular lattice with the nearest neighbor and next-nearest neighbor transfer integrals $t$ and $t^{\prime}$, while $t_d$ denotes the intradimer coupling.}
\label{fig:structrue}
\end{figure}

The present study is devoted to $\kappa$-(BEDT-TTF)$_2$\-Ag$_2$(CN)$_3$ ($\kappa$-AgCN, in short), a sibling compound to $\kappa$-CuCN.
These quasi-two dimensional organic charge-transfer salts possess a
rather similar crystal structure displayed in Fig.~\ref{fig:structrue}.
Layers of charge-donating organic BEDT-TTF (bis(ethylenedithio)-tetrathiafulvalene) mol\-e\-cules alternate with inorganic anion layers.
The (BEDT-TTF)$_2$ dimers form triangles with a high degree of geometrical frustration.
For $\kappa$-CuCN the ratio of the transfer integrals in the $bc$-plane $t'/t \approx 0.83$, whereas in  $\kappa$-AgCN is even closer to unity, $t'/t \approx 0.97$  \cite{Shimizu2016,Kandpal2009,Leylekian2020,Hartmann2019}.
Since electron-electron interactions are more pronounced in $\kappa$-AgCN compared to the Cu-analogue \cite{Pustogow2018}, the title compound constitutes a rare model of a strongly correlated Mott insulator with $S=\frac{1}{2}$ spins arranged on an almost perfect triangular lattice.
We have investigated the magnetic state of $\kappa$-AgCN by using broadband electron spin resonance spectroscopy down to $T = 0.56$~K.
In addition, the electrochemical stability of silver ions prevents the oxidation of nonmagnetic Ag(I) to paramagnetic Ag(II) \cite{Hiramatsu2017}; hence  $\kappa$-AgCN contains one potential source of disorder less than the $\kappa$-CuCN, where Cu$^{2+}$ impurity ions become a serious nuisance \cite{Padmalekha2015,Miksch2021,Pustogow2022}.
On the other hand, the difficulties in synthesizing $\kappa$-AgCN may likely come along with more inhomogeneities and disorder of other types. In the context of QSL, the knowledge of the ground state of $\kappa$-AgCN provides a crucial piece of information.

Unfortunately, the experimental basis for $\kappa$-AgCN is not as broad as it is for $\kappa$-CuCN \cite{Dressel2020,Pustogow2022}. Thermal expansion measurements do not reveal any sharp lattice anomaly down to $T = 1.5$~K, albeit there is a broad extremum of the expansion coefficient around 12~K with anomalous contraction up to 18-20~K \cite{Hartmann2019}. On the other hand, results of nuclear magnetic resonance (NMR) and specific heat were interpreted in terms of a gapless magnetic ground state \cite{Shimizu2016}. Recently the emergence of a low-entropy state has been inferred below around 11~K based on ac-transport measurements \cite{Rosslhuber2023}. Here, we present electron spin resonance (ESR) investigations of $\kappa$-AgCN single crystals. This method is known to be  crucial for any deeper understanding of the magnetic ground state \cite{Miksch2021}. We show that at high temperatures the magnetic response of $\kappa$-AgCN can be approximated by the susceptibility expected from a two-dimensional spin-$\frac{1}{2}$ Heisenberg antiferromagnet with an exchange interaction $J/k_B \approx 175$~K on a triangular lattice; most important, the rapid drop in spin susceptibility at low temperatures suggests a gapped magnetic ground state.

The single crystals of $\kappa$-AgCN were grown by electrochemical-oxidation method {\cite{Shimizu2016}} in different laboratories.
For sample characterization crystals were checked by resistivity and magnetic susceptibility measurements using standard methods. The crystals were oriented according to morphology
supplemented by infrared optical reflectivity \cite{Elsasser2012,Pustogow2018,Miksch2021}.
Most of our ESR spectra were collected in a Bruker X-band spectrometer ($\nu = 9.47$~GHz) equipped with a continuous flow  He-cryostat working in the temperature range down to $T = 1.8$~K when pumped. In addition, we have also employed a Q-band spectro\-meter (Bruker Elexsys 500) at 34 GHz operating down to 4~K. As a complementary method, superconducting coplanar resonators were used to obtain the ESR spectra at $\nu = 11.5$ and 18.7~GHz (K-band); for more details see Supplemental Materials \cite{SM}.


\begin{figure}
\centering
\includegraphics[width=0.7\columnwidth]{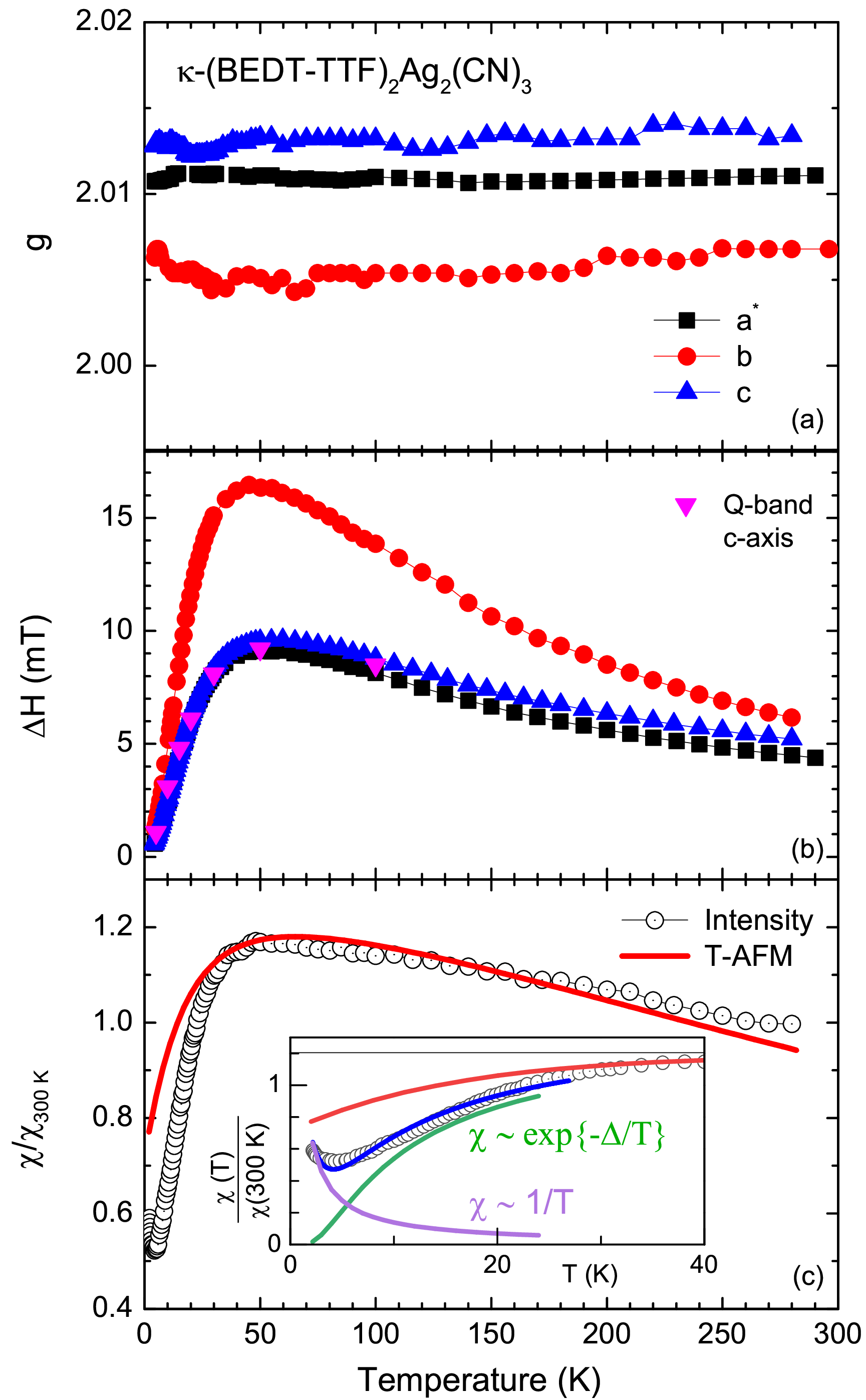}
\caption{Temperature dependence of (a) linewidth and (b) $g$-factor for $H_{\rm dc}$ parallel to the $a^*$, $b$,  and $c$ axes, as indicated, recorded in the X-band. The magenta down-pointing triangles in (b) corresponds to data taken at 34~GHz (Q-band) along the $c$-axis.
(c)~ESR intensity of $\kappa$-(BEDT-TTF)$_2$\-Ag$_2$(CN)$_3$ as a function of temperature.
Upon cooling a maximum occurs around 50~K before $\Delta H$ drops to less than half of its value. The red solid line corresponds to calculations of a two-dimensional $S=\frac{1}{2}$ Heisenberg antiferromagnet on a triangular lattice with a coupling of $J/k_B = 175$~K.
Below $T = 4$~K the spin susceptibility exhibits a small rise, as magnified in the inset. Besides the data (open circles) and the Heisenberg model (red line), we show a fit (blue line) of the low-$T$ susceptibility by a Curie contribution (purple line) and resultant spin-susceptibility of a gapped system (green line).}
\label{fig:ESR-parametes}
\end{figure}

At room temperature a single absorption line is observed in $\kappa$-AgCN along the three crystallographic axes, $b$, $c$ and $a^*$, which is perpendicular to the $bc$-plane \cite{SM}. While for $H_{\rm dc}\parallel a^{*}$ the ESR absorption can be modelled by a single Lorentzian lineshape, along the $b$ and $c$ directions the ESR lineshape is Dysonian in nature (Fig.~S2). The conductivity within the $bc$-plane is considerable at room-temperature, although $\kappa$-AgCN is a correlated insulator far away from the Mott metal-insulator boundary. This is in accord with dc, ac and optical conductivity measurements \cite{Shimizu2016,Pinteric2016,Pustogow2018,Rosslhuber2023}. For each temperature the ESR spectra are analyzed to obtain the center frequency
$h\nu = g \mu_0 \mu_B H_{\rm dc}$, the linewidth $\Delta H$ and the intensity of the signal $I_{\rm ESR}$.
In Fig.~\ref{fig:ESR-parametes}(a) $g(T)$ is plotted for each orientation $a^*$, $b$, and $c$;
no considerable temperature dependence is observed.
The linewidth $\Delta H$, on the other hand,  significantly varies with temperature, as presented in panel (b). As the temperature decreases, the lines become broader and $\Delta H(T)$ reaches a maximum around $T \approx 50$~K. Subsequently, $\Delta H(T)$ starts to decrease rapidly down to the lowest measured temperature of $T = 1.8$~K.
We also show the Q-band data recorded along the $c$-axis, which fall right on top of the X-band results. This is not surprising as the width is dominated by crystal-field effects.

\begin{figure}
	\centering
	\includegraphics[width=0.8\columnwidth]{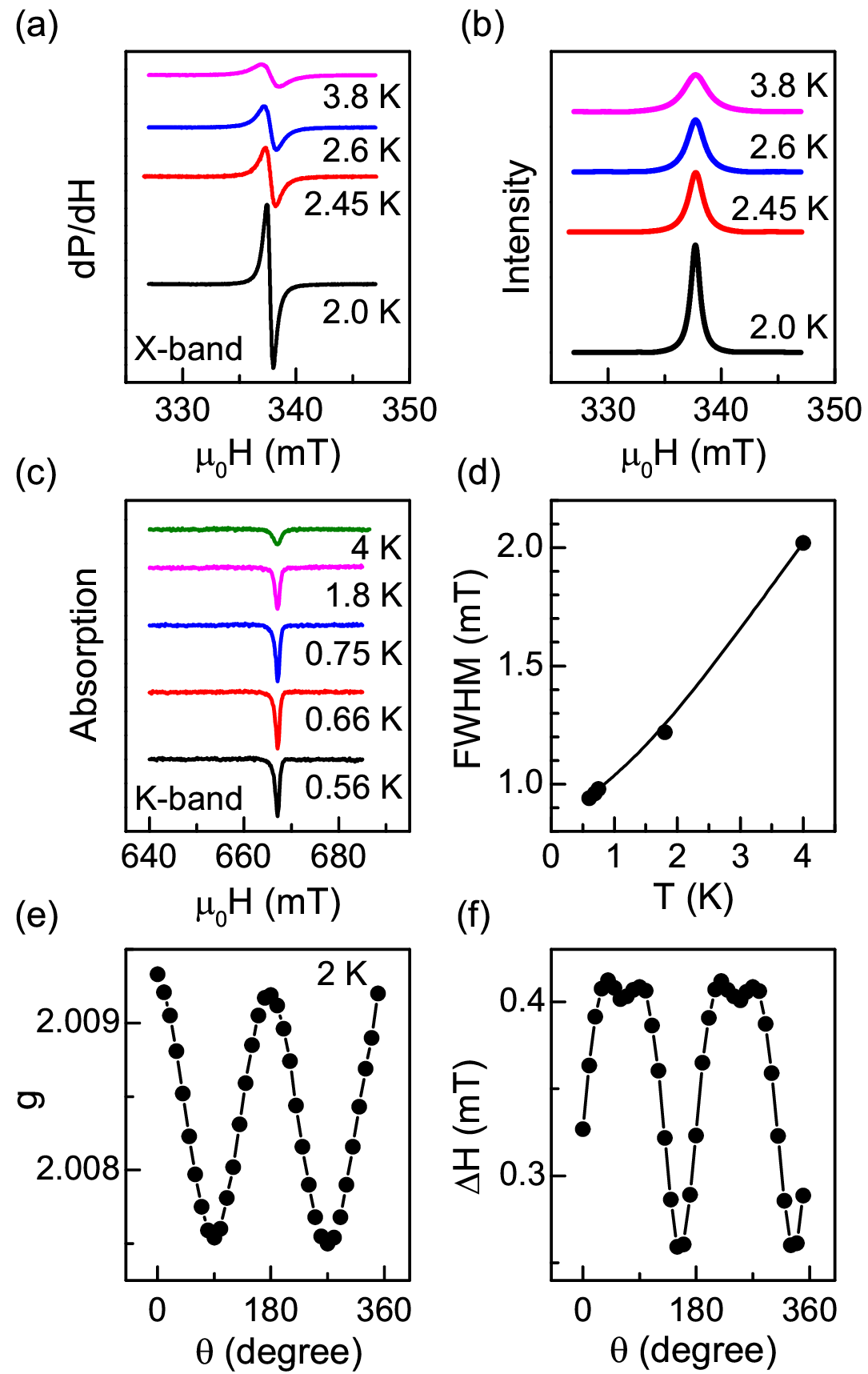}
	\caption{Low-temperature ESR investigations on $\kappa$-(BEDT-TTF)$_2$Ag$_2$(CN)$_3$.
		(a) X-band data and (b) integrated ESR spectra along the $b$-axis below $T = 4$~K.
		(c) K-band absorption measured by a parallel-plane resonator at $\nu = 18.7$~GHz down to $T=0.56$~K. (d) Temperature dependence of the corresponding full width at half maximum (FWHM). (e,f) Angular variation of the $g$-factor and linewidth $\Delta H$ recorded in the X-band at $T = 2$~K.}
	\label{fig:ESR2}
\end{figure}

In Fig.~\ref{fig:ESR-parametes}(c) the temperature dependence of the spin susceptibility $\chi_s(T) \propto I_{\rm ESR}$  normalized to its room-temperature value is plotted. As the temperature is reduced, the susceptibility increases slightly and exhibits a maximum around $T \approx 50$~K. At lower temperatures, it starts to decrease rapidly but does not vanish.
Below approximately 4~K, the spin susceptibility rises slightly resembling a Curie-tail.
Overall, the temperature dependence of $\chi_s(T)$ does not reveal any signature of a phase transition to a long-range magnetically ordered state and hence suggests that strong quantum fluctuations due to geometrical frustration of the triangular lattice suppress the antiferromagnetic ordering.
The solid red line  in panel (c) represents the theoretical curve for a two-dimensional
$S=\frac{1}{2}$ Heisenberg antiferromagnet on a triangular lattice (T-AFM)
with nearest-neighbor exchange interaction $J/k_B = 175$~K \cite{Elstner1993}.
In this model the magnetic susceptibility exhibits the maximum around $0.35 J$ and remains finite as $T \rightarrow 0$. The data of $\kappa$-AgCN, however, clearly deviate from the model for  $T < 30$~K. The experimental spin susceptibility decreases more rapidly at lower temperatures than predicted by the Heisenberg model, as displayed in the inset of Fig.~\ref{fig:ESR-parametes}(c).

In addition to the rapid reduction, $\chi_s(T)$ shows an upturn below $T \approx 4$~K.
Such an enhancement is frequently observed in QSL systems at low temperatures \cite{Padmalekha2015,Miksch2021}. Commonly this behavior is assigned to a Curie-type low-temperature $T^{-1}$ response of paramagnetic impurity spins. However, there are other scenarios to be considered here, too  \cite{Pohle2023}.
In a system with random disorder, one possible ground state is the random singlet phase of a spin liquid, where the susceptibility shows a power law $\chi_s(T) \propto T^{-\gamma}$, with $\gamma < 1$. Such phase is observed in systems with considerable amount of disorder like doped Si, YbMgGaO$_4$, $1T$-TaS$_2$ etc. {\cite{Furukawa2015,Kimchi2018,Pal2020,Pal2022}}.
The power-law dependence of $\chi_s(T)$ at low temperatures suggests the coexistence of random spin singlets and isolated localized magnetic moments in the host lattice.
The emergence of an inhomogeneous energy gap was also shown as a possible scenario in  materials, where spin singlets gradually appear with spatially varying energy gap \cite{Wang2021}.

In the low-temperature range, where the spin susceptibility of $\kappa$-AgCN decreases rapidly, also the linewidth $\Delta H$ monotonically drops very fast [Fig.~\ref{fig:ESR-parametes}(b)]. The ESR linewidth reflects the spin dynamics of the system, and hence it constitutes an extremely important parameter to understand the magnetic state. Such rapid decrease of ESR linewidth is often observed in one-dimensional spin-chain systems exhibiting a spin gap \cite{Chabre2005,Deisenhofer2006,Vasiliev2015}.  The decrease in $\Delta H(T)$ points towards the formation of spin singlets at lower temperatures. As a result, the relaxation time of the remaining orphan spins increases, yielding a reduction of the ESR linewidth.

Therefore, to better understand the rapid drop of the spin susceptibility at low $T$, we model $\chi_s(T)$ with a combination of two contributions: a Curie term and an exponential decay with a fixed energy gap $\Delta$. The fit shown in the inset of Fig.~\ref{fig:ESR-parametes}(c)
is obtained with $\Delta \approx 10$~K.
In addition, comparing $I_{\rm ESR}$ with the magnetic susceptibility known from standard SQUID magnetometry, we estimate an impurity concentration of $N \approx 0.002$ per unit cell, which matches quite well with the previous estimation \cite{Shimizu2016,Pustogow2020}. Not unexpected, we found considerable sample-to-sample variation of the Curie contribution (Fig.~S6 in \cite{SM}).
Interestingly, the first derivative of the susceptibility ${\rm d}\chi_s(T)/{\rm d}T$ indeed shows a broad feature around $T \approx 12$~K (Fig.~S4), indicative of a distribution of the energy gap with an average of $\Delta \approx 10$~K.

For a more detailed discussion, Fig.~\ref{fig:ESR2} shows the low-temperature results obtained on
$\kappa$-AgCN  along the $b$-axis at $\nu = 9.47$ and 18.7~GHz.
In both cases the ESR intensity and absorption reveals a simple single Lorentzian lineshape.
The K-band linewidth (FWHM) increases by more than a factor of 2 when the temperature increases from 0.56 to 4~K. This is consistent with the behavior of $\Delta H(T)$ observed in the X-band and Q-band at high temperatures, as displayed in Fig.~\ref{fig:ESR-parametes}(b).
Also the $g$-factor remains unchanged.
Most importantly, we do not see any additional ESR absorption. This is a strong indication that the spins, which give rise to the enhancement of the magnetic susceptibility, exhibit similar anisotropy as the host.
This behavior is quite different from the situation encountered in $\kappa$-CuCN, where two different ESR absorption lines are observed at low temperatures;
next to the main signal (intrinsic) a secondary signal is identified and associated with defect spins located in the vicinity of Cu$^{2+}$ impurities \cite{Miksch2021}.
The absence of such a secondary line in $\kappa$-AgCN is consistent with the absence Ag$^{2+}$.
While in $\kappa$-CuCN $\chi_s(T)$ decreases exponentially below $T^{\star}$, for $\kappa$-AgCN a roll-off sets in around 20 to 30~K. Nevertheless, the exponential behavior with an energy gap $\Delta \approx 10$~K gives a very good fit between 5 and 20~K.
The decrease in the linewidth discards the possibility of any long-range antiferromagnetic order down to $T=0.56$~K because this would lead to divergences of $\Delta H$ and FWHM.
Although we do not observe a clear signature of an additional absorption feature at low temperatures, the angular variation of the $g$-factor and $\Delta H$ displayed in Fig.~\ref{fig:ESR2}(e,f) provides crucial information.
The angular dependence of the linewidth infers the presence of two ESR lines at $T=2$~K;
these two lines possess similar anisotropy as indicated by the angular variation of the $g$-factor. Such correlations between the two ESR lines clearly underline the fact that the additional absorption is strongly correlated to the host spin system.

\begin{figure}[h]
	\centering
	\includegraphics[width=0.9\columnwidth]{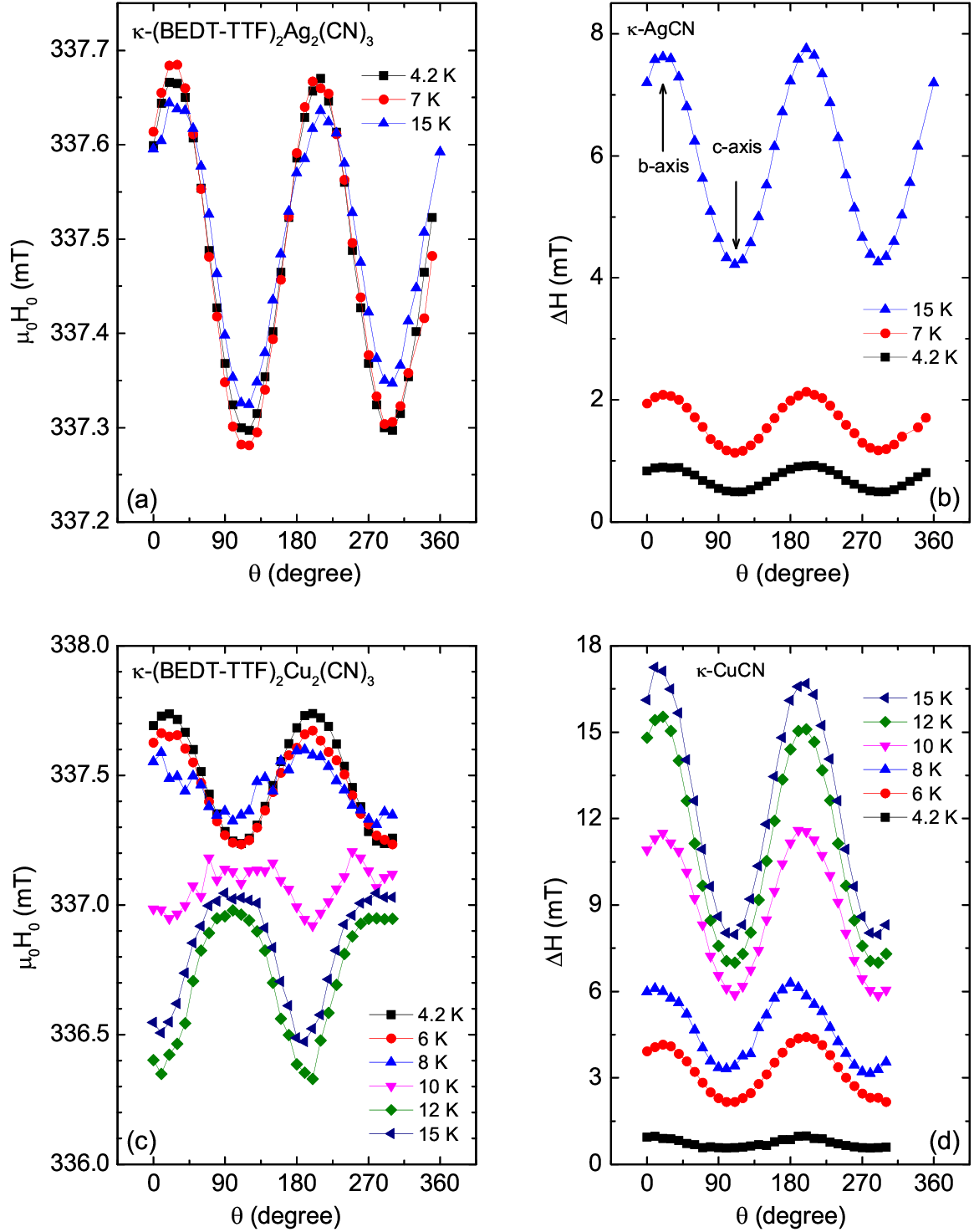}
	\caption{Angular dependence of $H_{\rm dc}$ and $\Delta H$ of (a,b) $\kappa$-AgCN and (c,d) $\kappa$-CuCN
		at several temperatures as indicated.
		These parameters are obtained from fits of the individual ESR spectra using a single ESR absorption of Lorentzian lineshape.
		For both the samples, the lines are broadest along the $b$-axis at all temperatures.
		However, we observe a reversal of the $g$-factor as the temperature is reduced.}
	\label{fig:ESR_3}
\end{figure}

At this point, we emphasize on the inversion of the $g$-factor anisotropy in $\kappa$-CuCN around $T^{\star}$ observed in the angular variation of the g-factor shown in Fig. \ref{fig:ESR_3}. While, the linewidth is unaltered between 4-15~K [Fig.~\ref{fig:ESR_3}(b,d)] in both compounds, the $g$-factor is reversed in $\kappa$-CuCN around $T^*$, as seen in Fig.~\ref{fig:ESR_3}(c). This is in line with the temperature dependence of the g-factor observed previously \cite{Miksch2021,SM}. Reversal of the g-factor has also been observed for the spin liquid candidates Herbertsmithite \cite{Zorko2017} and $\alpha$-RuCl$_3$ monolayers \cite{Yang2023}, which is proposed to be driven by lattice distortion. The title compound $\kappa$-AgCN however does not show a similar behavior: from Fig.~\ref{fig:ESR-parametes}(a) we see that the anisotropy of the $g$-factor remains throughout the entire temperature range.
This can be explained
by the similar $g$-factor anisotropy of spins that cause the enhancement of $\chi$ below 10~K and the host contribution. Another explanation may be the less pronounced lattice effects in $\kappa$-AgCN compared to $\kappa$-CuCN \cite{Manna2010,Pustogow2022,Matsuura2022},
albeit also for $\kappa$-AgCN there is a broad extremum in thermal expansion coefficient around 12~K \cite{Hartmann2019} coincident with low-entropy signatures below this temperature in pressure-dependent dielectric measurements \cite{Rosslhuber2023}.
Interestingly, this temperature scale agrees with the temperature range where the $^{13}$C NMR spectra starts broadening, stretched-exponential relaxation sets in and $T_1^{-1}$ steeply reduces towards a local minimum, below which a field-dependent maximum arises indicative of impurity spins \cite{Shimizu2016,Pustogow2020}.
Overall, the different situation compared to $\kappa$-CuCN may be the result of distinct lattice properties and the different origin of impurity spins.
On the one hand, it is rather likely that the unpaired electron spins on BEDT-TTF in the direct vicinity of a Cu$^{2+}$ impurity are strongly affected by a local lattice distortion
from the charged Cu impurity. On the other hand, such type of defects do not exist in $\kappa$-AgCN, hence the origin of unpaired impurity spins may lie in domain boundaries or other sources. We suggest targeted studies in this direction in future work.

On the theoretical side, although many models yield the presence of quantum-spin-liquid state, the precise nature of the ground state, whether it is gapped or not, remains inconclusive so far \cite{Jiang2023,Sherman2023,Szasz2020,Szasz2021,Iqbal2016,Shirakawa2017,Hu2019}. Indeed,
$S = \frac{1}{2}$ $J_1$-$J_2$ Heisenberg models on the triangular lattice predict gapped spin liquid states in presence of weak next-nearest-neighbor exchange \cite{Zhu2015,Hu2015,Chen2022}. In this respect, the strongly correlated compound $\kappa$-AgCN without significant lattice anomaly can serve as a model system.

In conclusion, we have investigated the magnetic state of the spin liquid candidate $\kappa$-AgCN  down to $T=0.56$~K using ESR spectroscopy at different frequencies.
Although the system with $S=\frac{1}{2}$ on a triangular lattice behaves as a two-dimensional Heisenberg antiferromagnet at high temperatures, the rapid drop of magnetic susceptibility and a weak Curie tail at low $T$ can be explained by assuming a gapped ground state with a small amount of impurity spins. An energy gap of $\Delta \approx 10$~K in magnetic susceptibility along with broad features observed in complementary experiments is taken as indication of a relevant characteristic temperature scale $T^* \approx  10$--20~K. The difference between the two prime spin-liquid candidate $\kappa$-CuCN and $\kappa$-AgCN might lie in the distinct involvement of the lattice degrees of freedom as well as other types and quantity of impurities.

We warmly thank M. Allain for checking the crystals produced in Angers; M. Girault is acknowledged for help with the preparation of the single crystals;  G. Untereiner provided technical help with the ESR experiments in Stuttgart.
We acknowledge the relevant discussions with K. Kanoda and R. K. Kremer.
This work was supported by the Deutsche Forschungsgemeinschaft (DFG) via DR228/68-1.


\begin{thebibliography}{56}%
\makeatletter
\providecommand \@ifxundefined [1]{%
 \@ifx{#1\undefined}
}%
\providecommand \@ifnum [1]{%
 \ifnum #1\expandafter \@firstoftwo
 \else \expandafter \@secondoftwo
 \fi
}%
\providecommand \@ifx [1]{%
 \ifx #1\expandafter \@firstoftwo
 \else \expandafter \@secondoftwo
 \fi
}%
\providecommand \natexlab [1]{#1}%
\providecommand \enquote  [1]{``#1''}%
\providecommand \bibnamefont  [1]{#1}%
\providecommand \bibfnamefont [1]{#1}%
\providecommand \citenamefont [1]{#1}%
\providecommand \href@noop [0]{\@secondoftwo}%
\providecommand \href [0]{\begingroup \@sanitize@url \@href}%
\providecommand \@href[1]{\@@startlink{#1}\@@href}%
\providecommand \@@href[1]{\endgroup#1\@@endlink}%
\providecommand \@sanitize@url [0]{\catcode `\\12\catcode `\$12\catcode
  `\&12\catcode `\#12\catcode `\^12\catcode `\_12\catcode `\%12\relax}%
\providecommand \@@startlink[1]{}%
\providecommand \@@endlink[0]{}%
\providecommand \url  [0]{\begingroup\@sanitize@url \@url }%
\providecommand \@url [1]{\endgroup\@href {#1}{\urlprefix }}%
\providecommand \urlprefix  [0]{URL }%
\providecommand \Eprint [0]{\href }%
\providecommand \doibase [0]{https://doi.org/}%
\providecommand \selectlanguage [0]{\@gobble}%
\providecommand \bibinfo  [0]{\@secondoftwo}%
\providecommand \bibfield  [0]{\@secondoftwo}%
\providecommand \translation [1]{[#1]}%
\providecommand \BibitemOpen [0]{}%
\providecommand \bibitemStop [0]{}%
\providecommand \bibitemNoStop [0]{.\EOS\space}%
\providecommand \EOS [0]{\spacefactor3000\relax}%
\providecommand \BibitemShut  [1]{\csname bibitem#1\endcsname}%
\let\auto@bib@innerbib\@empty
\bibitem [{\citenamefont {Zhou}\ \emph {et~al.}(2017)\citenamefont {Zhou},
  \citenamefont {Kanoda},\ and\ \citenamefont {Ng}}]{Zhou2017}%
  \BibitemOpen
  \bibfield  {author} {\bibinfo {author} {\bibfnamefont {Y.}~\bibnamefont
  {Zhou}}, \bibinfo {author} {\bibfnamefont {K.}~\bibnamefont {Kanoda}},\ and\
  \bibinfo {author} {\bibfnamefont {T.-K.}\ \bibnamefont {Ng}},\ }\bibfield
  {title} {\bibinfo {title} {Quantum spin liquid states},\ }\href
  {https://doi.org/10.1103/RevModPhys.89.025003} {\bibfield  {journal}
  {\bibinfo  {journal} {Rev. Mod. Phys.}\ }\textbf {\bibinfo {volume} {89}},\
  \bibinfo {pages} {025003} (\bibinfo {year} {2017})}\BibitemShut {NoStop}%
\bibitem [{\citenamefont {Broholm}\ \emph {et~al.}(2020)\citenamefont
  {Broholm}, \citenamefont {Cava}, \citenamefont {Kivelson}, \citenamefont
  {Nocera}, \citenamefont {Norman},\ and\ \citenamefont
  {Senthil}}]{Broholm2020}%
  \BibitemOpen
  \bibfield  {author} {\bibinfo {author} {\bibfnamefont {C.}~\bibnamefont
  {Broholm}}, \bibinfo {author} {\bibfnamefont {R.~J.}\ \bibnamefont {Cava}},
  \bibinfo {author} {\bibfnamefont {S.~A.}\ \bibnamefont {Kivelson}}, \bibinfo
  {author} {\bibfnamefont {D.~G.}\ \bibnamefont {Nocera}}, \bibinfo {author}
  {\bibfnamefont {M.~R.}\ \bibnamefont {Norman}},\ and\ \bibinfo {author}
  {\bibfnamefont {T.}~\bibnamefont {Senthil}},\ }\bibfield  {title} {\bibinfo
  {title} {{Quantum spin liquids}},\ }\href
  {https://doi.org/10.1126/science.aay0668} {\bibfield  {journal} {\bibinfo
  {journal} {Science}\ }\textbf {\bibinfo {volume} {367}},\ \bibinfo {pages}
  {eaay0668} (\bibinfo {year} {2020})}\BibitemShut {NoStop}%
\bibitem [{\citenamefont {Anderson}(1973)}]{Anderson1973}%
  \BibitemOpen
  \bibfield  {author} {\bibinfo {author} {\bibfnamefont {P.}~\bibnamefont
  {Anderson}},\ }\bibfield  {title} {\bibinfo {title} {Resonating valence
  bonds: A new kind of insulator?},\ }\href
  {https://doi.org/http://dx.doi.org/10.1016/0025-5408(73)90167-0} {\bibfield
  {journal} {\bibinfo  {journal} {Mater. Res. Bull.}\ }\textbf {\bibinfo
  {volume} {8}},\ \bibinfo {pages} {153} (\bibinfo {year} {1973})}\BibitemShut
  {NoStop}%
\bibitem [{\citenamefont {Fazekas}\ and\ \citenamefont
  {Anderson}(1974)}]{Fazekas1974}%
  \BibitemOpen
  \bibfield  {author} {\bibinfo {author} {\bibfnamefont {P.}~\bibnamefont
  {Fazekas}}\ and\ \bibinfo {author} {\bibfnamefont {P.~W.}\ \bibnamefont
  {Anderson}},\ }\bibfield  {title} {\bibinfo {title} {On the ground state
  properties of the anisotropic triangular antiferromagnet},\ }\href
  {https://doi.org/10.1080/14786439808206568} {\bibfield  {journal} {\bibinfo
  {journal} {Phil. Mag.}\ }\textbf {\bibinfo {volume} {30}},\ \bibinfo {pages}
  {423} (\bibinfo {year} {1974})}\BibitemShut {NoStop}%
\bibitem [{\citenamefont {Anderson}(1987)}]{Anderson1987}%
  \BibitemOpen
  \bibfield  {author} {\bibinfo {author} {\bibfnamefont {P.~W.}\ \bibnamefont
  {Anderson}},\ }\bibfield  {title} {\bibinfo {title} {The resonating valence
  bond state in {{La$_2$CuO$_4$}} and superconductivity},\ }\href
  {https://doi.org/10.1126/science.235.4793.1196} {\bibfield  {journal}
  {\bibinfo  {journal} {Science}\ }\textbf {\bibinfo {volume} {235}},\ \bibinfo
  {pages} {1196} (\bibinfo {year} {1987})}\BibitemShut {NoStop}%
\bibitem [{\citenamefont {Scheie}\ \emph {et~al.}(2023)\citenamefont {Scheie},
  \citenamefont {Ghioldi}, \citenamefont {Xing}, \citenamefont {Paddison},
  \citenamefont {Sherman}, \citenamefont {Dupont}, \citenamefont {Sanjeewa},
  \citenamefont {Lee}, \citenamefont {Woods}, \citenamefont {Abernathy},
  \citenamefont {Pajerowski}, \citenamefont {Williams}, \citenamefont {Zhang},
  \citenamefont {Manuel}, \citenamefont {Trumper}, \citenamefont {Pemmaraju},
  \citenamefont {Sefat}, \citenamefont {Parker}, \citenamefont {Devereaux},
  \citenamefont {Movshovich}, \citenamefont {Moore}, \citenamefont {Batista},\
  and\ \citenamefont {Tennant}}]{Scheie2023}%
  \BibitemOpen
  \bibfield  {author} {\bibinfo {author} {\bibfnamefont {A.~O.}\ \bibnamefont
  {Scheie}}, \bibinfo {author} {\bibfnamefont {E.~A.}\ \bibnamefont {Ghioldi}},
  \bibinfo {author} {\bibfnamefont {J.}~\bibnamefont {Xing}}, \bibinfo {author}
  {\bibfnamefont {J.~A.~M.}\ \bibnamefont {Paddison}}, \bibinfo {author}
  {\bibfnamefont {N.~E.}\ \bibnamefont {Sherman}}, \bibinfo {author}
  {\bibfnamefont {M.}~\bibnamefont {Dupont}}, \bibinfo {author} {\bibfnamefont
  {L.~D.}\ \bibnamefont {Sanjeewa}}, \bibinfo {author} {\bibfnamefont
  {S.}~\bibnamefont {Lee}}, \bibinfo {author} {\bibfnamefont {A.~J.}\
  \bibnamefont {Woods}}, \bibinfo {author} {\bibfnamefont {D.}~\bibnamefont
  {Abernathy}}, \bibinfo {author} {\bibfnamefont {D.~M.}\ \bibnamefont
  {Pajerowski}}, \bibinfo {author} {\bibfnamefont {T.~J.}\ \bibnamefont
  {Williams}}, \bibinfo {author} {\bibfnamefont {S.-S.}\ \bibnamefont {Zhang}},
  \bibinfo {author} {\bibfnamefont {L.~O.}\ \bibnamefont {Manuel}}, \bibinfo
  {author} {\bibfnamefont {A.~E.}\ \bibnamefont {Trumper}}, \bibinfo {author}
  {\bibfnamefont {C.~D.}\ \bibnamefont {Pemmaraju}}, \bibinfo {author}
  {\bibfnamefont {A.~S.}\ \bibnamefont {Sefat}}, \bibinfo {author}
  {\bibfnamefont {D.~S.}\ \bibnamefont {Parker}}, \bibinfo {author}
  {\bibfnamefont {T.~P.}\ \bibnamefont {Devereaux}}, \bibinfo {author}
  {\bibfnamefont {R.}~\bibnamefont {Movshovich}}, \bibinfo {author}
  {\bibfnamefont {J.~E.}\ \bibnamefont {Moore}}, \bibinfo {author}
  {\bibfnamefont {C.~D.}\ \bibnamefont {Batista}},\ and\ \bibinfo {author}
  {\bibfnamefont {D.~A.}\ \bibnamefont {Tennant}},\ }\bibfield  {title}
  {\bibinfo {title} {Proximate spin liquid and fractionalization in the
  triangular antiferromagnet {{KYbSe$_2$}}},\ }\href
  {https://doi.org/10.1038/s41567-023-02259-1} {\bibfield  {journal} {\bibinfo
  {journal} {Nat. Phys.}\ }\textbf {\bibinfo {volume} {20}},\ \bibinfo {pages}
  {74} (\bibinfo {year} {2023})}\BibitemShut {NoStop}%
\bibitem [{\citenamefont {Shirata}\ \emph {et~al.}(2012)\citenamefont
  {Shirata}, \citenamefont {Tanaka}, \citenamefont {Matsuo},\ and\
  \citenamefont {Kindo}}]{Shirata2012}%
  \BibitemOpen
  \bibfield  {author} {\bibinfo {author} {\bibfnamefont {Y.}~\bibnamefont
  {Shirata}}, \bibinfo {author} {\bibfnamefont {H.}~\bibnamefont {Tanaka}},
  \bibinfo {author} {\bibfnamefont {A.}~\bibnamefont {Matsuo}},\ and\ \bibinfo
  {author} {\bibfnamefont {K.}~\bibnamefont {Kindo}},\ }\bibfield  {title}
  {\bibinfo {title} {Experimental realization of a spin-$1/2$
  triangular-lattice heisenberg antiferromagnet},\ }\href
  {https://doi.org/10.1103/PhysRevLett.108.057205} {\bibfield  {journal}
  {\bibinfo  {journal} {Phys. Rev. Lett.}\ }\textbf {\bibinfo {volume} {108}},\
  \bibinfo {pages} {057205} (\bibinfo {year} {2012})}\BibitemShut {NoStop}%
\bibitem [{\citenamefont {Susuki}\ \emph {et~al.}(2013)\citenamefont {Susuki},
  \citenamefont {Kurita}, \citenamefont {Tanaka}, \citenamefont {Nojiri},
  \citenamefont {Matsuo}, \citenamefont {Kindo},\ and\ \citenamefont
  {Tanaka}}]{Susuki2013}%
  \BibitemOpen
  \bibfield  {author} {\bibinfo {author} {\bibfnamefont {T.}~\bibnamefont
  {Susuki}}, \bibinfo {author} {\bibfnamefont {N.}~\bibnamefont {Kurita}},
  \bibinfo {author} {\bibfnamefont {T.}~\bibnamefont {Tanaka}}, \bibinfo
  {author} {\bibfnamefont {H.}~\bibnamefont {Nojiri}}, \bibinfo {author}
  {\bibfnamefont {A.}~\bibnamefont {Matsuo}}, \bibinfo {author} {\bibfnamefont
  {K.}~\bibnamefont {Kindo}},\ and\ \bibinfo {author} {\bibfnamefont
  {H.}~\bibnamefont {Tanaka}},\ }\bibfield  {title} {\bibinfo {title}
  {Magnetization process and collective excitations in the $s\mathbf{=}1/2$
  triangular-lattice {{Heisenberg}} antiferromagnet
  {{Ba$_{3}$CoSb$_{2}$O$_{9}$}}},\ }\href
  {https://doi.org/10.1103/PhysRevLett.110.267201} {\bibfield  {journal}
  {\bibinfo  {journal} {Phys. Rev. Lett.}\ }\textbf {\bibinfo {volume} {110}},\
  \bibinfo {pages} {267201} (\bibinfo {year} {2013})}\BibitemShut {NoStop}%
\bibitem [{\citenamefont {Khuntia}\ \emph {et~al.}(2020)\citenamefont
  {Khuntia}, \citenamefont {Velazquez}, \citenamefont {Barth{\'{e}}lemy},
  \citenamefont {Bert}, \citenamefont {Kermarrec}, \citenamefont {Legros},
  \citenamefont {Bernu}, \citenamefont {Messio}, \citenamefont {Zorko},\ and\
  \citenamefont {Mendels}}]{Khuntia2020}%
  \BibitemOpen
  \bibfield  {author} {\bibinfo {author} {\bibfnamefont {P.}~\bibnamefont
  {Khuntia}}, \bibinfo {author} {\bibfnamefont {M.}~\bibnamefont {Velazquez}},
  \bibinfo {author} {\bibfnamefont {Q.}~\bibnamefont {Barth{\'{e}}lemy}},
  \bibinfo {author} {\bibfnamefont {F.}~\bibnamefont {Bert}}, \bibinfo {author}
  {\bibfnamefont {E.}~\bibnamefont {Kermarrec}}, \bibinfo {author}
  {\bibfnamefont {A.}~\bibnamefont {Legros}}, \bibinfo {author} {\bibfnamefont
  {B.}~\bibnamefont {Bernu}}, \bibinfo {author} {\bibfnamefont
  {L.}~\bibnamefont {Messio}}, \bibinfo {author} {\bibfnamefont
  {A.}~\bibnamefont {Zorko}},\ and\ \bibinfo {author} {\bibfnamefont
  {P.}~\bibnamefont {Mendels}},\ }\bibfield  {title} {\bibinfo {title} {Gapless
  ground state in the archetypal quantum kagome antiferromagnet
  {{ZnCu$_3$(OH)$_6$Cl$_2$}}},\ }\href
  {https://doi.org/10.1038/s41567-020-0792-1} {\bibfield  {journal} {\bibinfo
  {journal} {Nat. Phys.}\ }\textbf {\bibinfo {volume} {16}},\ \bibinfo {pages}
  {469} (\bibinfo {year} {2020})}\BibitemShut {NoStop}%
\bibitem [{\citenamefont {Shimizu}\ \emph {et~al.}(2003)\citenamefont
  {Shimizu}, \citenamefont {Miyagawa}, \citenamefont {Kanoda}, \citenamefont
  {Maesato},\ and\ \citenamefont {Saito}}]{Shimizu2003}%
  \BibitemOpen
  \bibfield  {author} {\bibinfo {author} {\bibfnamefont {Y.}~\bibnamefont
  {Shimizu}}, \bibinfo {author} {\bibfnamefont {K.}~\bibnamefont {Miyagawa}},
  \bibinfo {author} {\bibfnamefont {K.}~\bibnamefont {Kanoda}}, \bibinfo
  {author} {\bibfnamefont {M.}~\bibnamefont {Maesato}},\ and\ \bibinfo {author}
  {\bibfnamefont {G.}~\bibnamefont {Saito}},\ }\bibfield  {title} {\bibinfo
  {title} {Spin liquid state in an organic {{Mott}} insulator with a triangular
  lattice},\ }\href {https://doi.org/doi/10.1103/PhysRevLett.91.107001}
  {\bibfield  {journal} {\bibinfo  {journal} {Phys. Rev. Lett.}\ }\textbf
  {\bibinfo {volume} {91}},\ \bibinfo {pages} {107001} (\bibinfo {year}
  {2003})}\BibitemShut {NoStop}%
\bibitem [{\citenamefont {Itou}\ \emph {et~al.}(2010)\citenamefont {Itou},
  \citenamefont {Oyamada}, \citenamefont {Maegawa},\ and\ \citenamefont
  {Kato}}]{Itou2010}%
  \BibitemOpen
  \bibfield  {author} {\bibinfo {author} {\bibfnamefont {T.}~\bibnamefont
  {Itou}}, \bibinfo {author} {\bibfnamefont {A.}~\bibnamefont {Oyamada}},
  \bibinfo {author} {\bibfnamefont {S.}~\bibnamefont {Maegawa}},\ and\ \bibinfo
  {author} {\bibfnamefont {R.}~\bibnamefont {Kato}},\ }\bibfield  {title}
  {\bibinfo {title} {{Instability of a quantum spin liquid in an organic
  triangular-lattice antiferromagnet}},\ }\href
  {http://dx.doi.org/10.1038/nphys1715} {\bibfield  {journal} {\bibinfo
  {journal} {Nat. Phys.}\ }\textbf {\bibinfo {volume} {6}},\ \bibinfo {pages}
  {673} (\bibinfo {year} {2010})}\BibitemShut {NoStop}%
\bibitem [{\citenamefont {Shimizu}\ \emph {et~al.}(2016)\citenamefont
  {Shimizu}, \citenamefont {Hiramatsu}, \citenamefont {Maesato}, \citenamefont
  {Otsuka}, \citenamefont {Yamochi}, \citenamefont {Ono}, \citenamefont {Itoh},
  \citenamefont {Yoshida}, \citenamefont {Takigawa}, \citenamefont {Yoshida},\
  and\ \citenamefont {Saito}}]{Shimizu2016}%
  \BibitemOpen
  \bibfield  {author} {\bibinfo {author} {\bibfnamefont {Y.}~\bibnamefont
  {Shimizu}}, \bibinfo {author} {\bibfnamefont {T.}~\bibnamefont {Hiramatsu}},
  \bibinfo {author} {\bibfnamefont {M.}~\bibnamefont {Maesato}}, \bibinfo
  {author} {\bibfnamefont {A.}~\bibnamefont {Otsuka}}, \bibinfo {author}
  {\bibfnamefont {H.}~\bibnamefont {Yamochi}}, \bibinfo {author} {\bibfnamefont
  {A.}~\bibnamefont {Ono}}, \bibinfo {author} {\bibfnamefont {M.}~\bibnamefont
  {Itoh}}, \bibinfo {author} {\bibfnamefont {M.}~\bibnamefont {Yoshida}},
  \bibinfo {author} {\bibfnamefont {M.}~\bibnamefont {Takigawa}}, \bibinfo
  {author} {\bibfnamefont {Y.}~\bibnamefont {Yoshida}},\ and\ \bibinfo {author}
  {\bibfnamefont {G.}~\bibnamefont {Saito}},\ }\bibfield  {title} {\bibinfo
  {title} {Pressure-tuned exchange coupling of a quantum spin liquid in the
  molecular triangular lattice $\kappa$-{{(ET)$_{2}$Ag$_{2}$(CN)$_{3}$}}},\
  }\href {https://doi.org/doi/10.1103/PhysRevLett.117.107203} {\bibfield
  {journal} {\bibinfo  {journal} {Phys. Rev. Lett.}\ }\textbf {\bibinfo
  {volume} {117}},\ \bibinfo {pages} {107203} (\bibinfo {year}
  {2016})}\BibitemShut {NoStop}%
\bibitem [{\citenamefont {Dressel}\ and\ \citenamefont
  {Tomi{\'{c}}}(2020)}]{Dressel2020}%
  \BibitemOpen
  \bibfield  {author} {\bibinfo {author} {\bibfnamefont {M.}~\bibnamefont
  {Dressel}}\ and\ \bibinfo {author} {\bibfnamefont {S.}~\bibnamefont
  {Tomi{\'{c}}}},\ }\bibfield  {title} {\bibinfo {title} {{Molecular quantum
  materials: electronic phases and charge dynamics in two-dimensional organic
  solids}},\ }\href {https://doi.org/10.1080/00018732.2020.1837833} {\bibfield
  {journal} {\bibinfo  {journal} {Adv. Phys.}\ }\textbf {\bibinfo {volume}
  {69}},\ \bibinfo {pages} {1} (\bibinfo {year} {2020})}\BibitemShut {NoStop}%
\bibitem [{\citenamefont {Manna}\ \emph {et~al.}(2010)\citenamefont {Manna},
  \citenamefont {{de~Souza}}, \citenamefont {Br\"uhl}, \citenamefont
  {Schlueter},\ and\ \citenamefont {Lang}}]{Manna2010}%
  \BibitemOpen
  \bibfield  {author} {\bibinfo {author} {\bibfnamefont {R.~S.}\ \bibnamefont
  {Manna}}, \bibinfo {author} {\bibfnamefont {M.}~\bibnamefont {{de~Souza}}},
  \bibinfo {author} {\bibfnamefont {A.}~\bibnamefont {Br\"uhl}}, \bibinfo
  {author} {\bibfnamefont {J.~A.}\ \bibnamefont {Schlueter}},\ and\ \bibinfo
  {author} {\bibfnamefont {M.}~\bibnamefont {Lang}},\ }\bibfield  {title}
  {\bibinfo {title} {Lattice effects and entropy release at the low-temperature
  phase transition in the spin-liquid candidate
  {{$\kappa$-(BEDT-TTF)$_{2}$Cu$_{2}$(CN)$_{3}$}}},\ }\href
  {https://doi.org/10.1103/PhysRevLett.104.016403} {\bibfield  {journal}
  {\bibinfo  {journal} {Phys. Rev. Lett.}\ }\textbf {\bibinfo {volume} {104}},\
  \bibinfo {eid} {016403} (\bibinfo {year} {2010})}\BibitemShut {NoStop}%
\bibitem [{\citenamefont {Poirier}\ \emph {et~al.}(2014)\citenamefont
  {Poirier}, \citenamefont {{{de~Lafontaine}}}, \citenamefont {Miyagawa},
  \citenamefont {Kanoda},\ and\ \citenamefont {Shimizu}}]{Poirier2014}%
  \BibitemOpen
  \bibfield  {author} {\bibinfo {author} {\bibfnamefont {M.}~\bibnamefont
  {Poirier}}, \bibinfo {author} {\bibfnamefont {M.}~\bibnamefont
  {{{de~Lafontaine}}}}, \bibinfo {author} {\bibfnamefont {K.}~\bibnamefont
  {Miyagawa}}, \bibinfo {author} {\bibfnamefont {K.}~\bibnamefont {Kanoda}},\
  and\ \bibinfo {author} {\bibfnamefont {Y.}~\bibnamefont {Shimizu}},\
  }\bibfield  {title} {\bibinfo {title} {Ultrasonic investigation of the
  transition at {{6 K}} in the spin-liquid candidate
  $\kappa$-{{(BEDT-TTF)$_{2}$Cu$_{2}$(CN)$_{3}$}}},\ }\href
  {https://doi.org/10.1103/PhysRevB.89.045138} {\bibfield  {journal} {\bibinfo
  {journal} {Phys. Rev. B}\ }\textbf {\bibinfo {volume} {89}},\ \bibinfo {eid}
  {045138} (\bibinfo {year} {2014})}\BibitemShut {NoStop}%
\bibitem [{\citenamefont {Padmalekha}\ \emph {et~al.}(2015)\citenamefont
  {Padmalekha}, \citenamefont {Blankenhorn}, \citenamefont {Ivek},
  \citenamefont {Bogani}, \citenamefont {Schlueter},\ and\ \citenamefont
  {Dressel}}]{Padmalekha2015}%
  \BibitemOpen
  \bibfield  {author} {\bibinfo {author} {\bibfnamefont {K.~G.}\ \bibnamefont
  {Padmalekha}}, \bibinfo {author} {\bibfnamefont {M.}~\bibnamefont
  {Blankenhorn}}, \bibinfo {author} {\bibfnamefont {T.}~\bibnamefont {Ivek}},
  \bibinfo {author} {\bibfnamefont {L.}~\bibnamefont {Bogani}}, \bibinfo
  {author} {\bibfnamefont {J.~A.}\ \bibnamefont {Schlueter}},\ and\ \bibinfo
  {author} {\bibfnamefont {M.}~\bibnamefont {Dressel}},\ }\bibfield  {title}
  {\bibinfo {title} {{{ESR}} studies on the spin-liquid candidate
  $\kappa$-{{(BEDT-TTF)$_2$Cu$_2$(CN)$_3$}}: Anomalous response below {{$T
  =8$~K}}},\ }\href {https://doi.org/10.1016/j.physb.2014.11.073} {\bibfield
  {journal} {\bibinfo  {journal} {Physica B}\ }\textbf {\bibinfo {volume}
  {460}},\ \bibinfo {pages} {211} (\bibinfo {year} {2015})}\BibitemShut
  {NoStop}%
\bibitem [{\citenamefont {Miksch}\ \emph {et~al.}(2021)\citenamefont {Miksch},
  \citenamefont {Pustogow}, \citenamefont {{Javaheri Rahim}}, \citenamefont
  {Bardin}, \citenamefont {Kanoda}, \citenamefont {Schlueter}, \citenamefont
  {H{\"{u}}bner}, \citenamefont {Scheffler},\ and\ \citenamefont
  {Dressel}}]{Miksch2021}%
  \BibitemOpen
  \bibfield  {author} {\bibinfo {author} {\bibfnamefont {B.}~\bibnamefont
  {Miksch}}, \bibinfo {author} {\bibfnamefont {A.}~\bibnamefont {Pustogow}},
  \bibinfo {author} {\bibfnamefont {M.}~\bibnamefont {{Javaheri Rahim}}},
  \bibinfo {author} {\bibfnamefont {A.~A.}\ \bibnamefont {Bardin}}, \bibinfo
  {author} {\bibfnamefont {K.}~\bibnamefont {Kanoda}}, \bibinfo {author}
  {\bibfnamefont {J.~A.}\ \bibnamefont {Schlueter}}, \bibinfo {author}
  {\bibfnamefont {R.}~\bibnamefont {H{\"{u}}bner}}, \bibinfo {author}
  {\bibfnamefont {M.}~\bibnamefont {Scheffler}},\ and\ \bibinfo {author}
  {\bibfnamefont {M.}~\bibnamefont {Dressel}},\ }\bibfield  {title} {\bibinfo
  {title} {Gapped magnetic ground state in quantum spin liquid candidate
  {{$\kappa$-(BEDT-TTF)$_2$Cu$_2$(CN)$_3$}}},\ }\href
  {https://doi.org/10.1126/science.abc6363} {\bibfield  {journal} {\bibinfo
  {journal} {Science}\ }\textbf {\bibinfo {volume} {372}},\ \bibinfo {pages}
  {276} (\bibinfo {year} {2021})}\BibitemShut {NoStop}%
\bibitem [{\citenamefont {Pustogow}(2022)}]{Pustogow2022}%
  \BibitemOpen
  \bibfield  {author} {\bibinfo {author} {\bibfnamefont {A.}~\bibnamefont
  {Pustogow}},\ }\bibfield  {title} {\bibinfo {title} {Thirty-year anniversary
  of $\kappa$-{{(BEDT-TTF)$_2$Cu$_2$(CN)$_3$}}: Reconciling the spin gap in a
  spin-liquid candidate},\ }\href {https://doi.org/10.3390/solids3010007}
  {\bibfield  {journal} {\bibinfo  {journal} {Solids}\ }\textbf {\bibinfo
  {volume} {3}},\ \bibinfo {pages} {93} (\bibinfo {year} {2022})}\BibitemShut
  {NoStop}%
\bibitem [{\citenamefont {Pustogow}\ \emph {et~al.}(2018)\citenamefont
  {Pustogow}, \citenamefont {Bories}, \citenamefont {L{\"{o}}hle},
  \citenamefont {R{\"{o}}sslhuber}, \citenamefont {Zhukova}, \citenamefont
  {Gorshunov}, \citenamefont {Tomi{\'{c}}}, \citenamefont {Schlueter},
  \citenamefont {H{\"{u}}bner}, \citenamefont {Hiramatsu}, \citenamefont
  {Yoshida}, \citenamefont {Saito}, \citenamefont {Kato}, \citenamefont {Lee},
  \citenamefont {Dobrosavljevi{\'{c}}}, \citenamefont {Fratini},\ and\
  \citenamefont {Dressel}}]{Pustogow2018}%
  \BibitemOpen
  \bibfield  {author} {\bibinfo {author} {\bibfnamefont {A.}~\bibnamefont
  {Pustogow}}, \bibinfo {author} {\bibfnamefont {M.}~\bibnamefont {Bories}},
  \bibinfo {author} {\bibfnamefont {A.}~\bibnamefont {L{\"{o}}hle}}, \bibinfo
  {author} {\bibfnamefont {R.}~\bibnamefont {R{\"{o}}sslhuber}}, \bibinfo
  {author} {\bibfnamefont {E.}~\bibnamefont {Zhukova}}, \bibinfo {author}
  {\bibfnamefont {B.}~\bibnamefont {Gorshunov}}, \bibinfo {author}
  {\bibfnamefont {S.}~\bibnamefont {Tomi{\'{c}}}}, \bibinfo {author}
  {\bibfnamefont {J.~A.}\ \bibnamefont {Schlueter}}, \bibinfo {author}
  {\bibfnamefont {R.}~\bibnamefont {H{\"{u}}bner}}, \bibinfo {author}
  {\bibfnamefont {T.}~\bibnamefont {Hiramatsu}}, \bibinfo {author}
  {\bibfnamefont {Y.}~\bibnamefont {Yoshida}}, \bibinfo {author} {\bibfnamefont
  {G.}~\bibnamefont {Saito}}, \bibinfo {author} {\bibfnamefont
  {R.}~\bibnamefont {Kato}}, \bibinfo {author} {\bibfnamefont {T.-H.}\
  \bibnamefont {Lee}}, \bibinfo {author} {\bibfnamefont {V.}~\bibnamefont
  {Dobrosavljevi{\'{c}}}}, \bibinfo {author} {\bibfnamefont {S.}~\bibnamefont
  {Fratini}},\ and\ \bibinfo {author} {\bibfnamefont {M.}~\bibnamefont
  {Dressel}},\ }\bibfield  {title} {\bibinfo {title} {{Quantum spin liquids
  unveil the genuine Mott state}},\ }\href
  {https://doi.org/10.1038/s41563-018-0140-3} {\bibfield  {journal} {\bibinfo
  {journal} {Nat. Mater.}\ }\textbf {\bibinfo {volume} {17}},\ \bibinfo {pages}
  {773} (\bibinfo {year} {2018})}\BibitemShut {NoStop}%
\bibitem [{\citenamefont {Pustogow}\ \emph {et~al.}(2023)\citenamefont
  {Pustogow}, \citenamefont {Kawasugi}, \citenamefont {Sakurakoji},\ and\
  \citenamefont {Tajima}}]{Pustogow2023}%
  \BibitemOpen
  \bibfield  {author} {\bibinfo {author} {\bibfnamefont {A.}~\bibnamefont
  {Pustogow}}, \bibinfo {author} {\bibfnamefont {Y.}~\bibnamefont {Kawasugi}},
  \bibinfo {author} {\bibfnamefont {H.}~\bibnamefont {Sakurakoji}},\ and\
  \bibinfo {author} {\bibfnamefont {N.}~\bibnamefont {Tajima}},\ }\bibfield
  {title} {\bibinfo {title} {Chasing the spin gap through the phase diagram of
  a frustrated {{Mott}} insulator},\ }\href
  {https://doi.org/10.1038/s41467-023-37491-z} {\bibfield  {journal} {\bibinfo
  {journal} {Nat. Commun.}\ }\textbf {\bibinfo {volume} {14}},\ \bibinfo
  {pages} {1960} (\bibinfo {year} {2023})}\BibitemShut {NoStop}%
\bibitem [{\citenamefont {Kawasugi}\ \emph {et~al.}(2023)\citenamefont
  {Kawasugi}, \citenamefont {Yamazaki}, \citenamefont {Pustogow},\ and\
  \citenamefont {Tajima}}]{Kawasugi2023}%
  \BibitemOpen
  \bibfield  {author} {\bibinfo {author} {\bibfnamefont {Y.}~\bibnamefont
  {Kawasugi}}, \bibinfo {author} {\bibfnamefont {S.}~\bibnamefont {Yamazaki}},
  \bibinfo {author} {\bibfnamefont {A.}~\bibnamefont {Pustogow}},\ and\
  \bibinfo {author} {\bibfnamefont {N.}~\bibnamefont {Tajima}},\ }\bibfield
  {title} {\bibinfo {title} {Negative magnetoresistance near the mott
  metal–insulator transition in the quantum spin liquid candidate
  $\kappa$-{{(BEDT-TTF)$_2$Cu$_2$(CN)$_3$}}},\ }\href
  {https://doi.org/10.7566/JPSJ.92.065001} {\bibfield  {journal} {\bibinfo
  {journal} {J. Phys. Soc. Jpn.}\ }\textbf {\bibinfo {volume} {92}},\ \bibinfo
  {pages} {065001} (\bibinfo {year} {2023})}\BibitemShut {NoStop}%
\bibitem [{\citenamefont {Riedl}\ \emph {et~al.}(2019)\citenamefont {Riedl},
  \citenamefont {Valent{\'{i}}},\ and\ \citenamefont {Winter}}]{Riedl2019}%
  \BibitemOpen
  \bibfield  {author} {\bibinfo {author} {\bibfnamefont {K.}~\bibnamefont
  {Riedl}}, \bibinfo {author} {\bibfnamefont {R.}~\bibnamefont
  {Valent{\'{i}}}},\ and\ \bibinfo {author} {\bibfnamefont {S.~M.}\
  \bibnamefont {Winter}},\ }\bibfield  {title} {\bibinfo {title} {Critical spin
  liquid versus valence-bond glass in a triangular-lattice organic
  antiferromagnet},\ }\href {https://doi.org/10.1038/s41467-019-10604-3}
  {\bibfield  {journal} {\bibinfo  {journal} {Nat. Commun.}\ }\textbf {\bibinfo
  {volume} {10}},\ \bibinfo {pages} {2561} (\bibinfo {year}
  {2019})}\BibitemShut {NoStop}%
\bibitem [{\citenamefont {Matsuura}\ \emph {et~al.}(2022)\citenamefont
  {Matsuura}, \citenamefont {Sasaki}, \citenamefont {Naka}, \citenamefont
  {M\"uller}, \citenamefont {Stockert}, \citenamefont {Piovano}, \citenamefont
  {Yoneyama},\ and\ \citenamefont {Lang}}]{Matsuura2022}%
  \BibitemOpen
  \bibfield  {author} {\bibinfo {author} {\bibfnamefont {M.}~\bibnamefont
  {Matsuura}}, \bibinfo {author} {\bibfnamefont {T.}~\bibnamefont {Sasaki}},
  \bibinfo {author} {\bibfnamefont {M.}~\bibnamefont {Naka}}, \bibinfo {author}
  {\bibfnamefont {J.}~\bibnamefont {M\"uller}}, \bibinfo {author}
  {\bibfnamefont {O.}~\bibnamefont {Stockert}}, \bibinfo {author}
  {\bibfnamefont {A.}~\bibnamefont {Piovano}}, \bibinfo {author} {\bibfnamefont
  {N.}~\bibnamefont {Yoneyama}},\ and\ \bibinfo {author} {\bibfnamefont
  {M.}~\bibnamefont {Lang}},\ }\bibfield  {title} {\bibinfo {title} {Phonon
  renormalization effects accompanying the {{6 K}} anomaly in the quantum spin
  liquid candidate $\kappa$-{{(BEDT-TTF)$_{2}$Cu$_{2}$(CN)$_{3}$}}},\ }\href
  {https://doi.org/10.1103/PhysRevResearch.4.L042047} {\bibfield  {journal}
  {\bibinfo  {journal} {Phys. Rev. Res.}\ }\textbf {\bibinfo {volume} {4}},\
  \bibinfo {pages} {L042047} (\bibinfo {year} {2022})}\BibitemShut {NoStop}%
\bibitem [{\citenamefont {Riedl}\ \emph {et~al.}(2022)\citenamefont {Riedl},
  \citenamefont {Gati},\ and\ \citenamefont {Valent{'i}}}]{Riedl2022}%
  \BibitemOpen
  \bibfield  {author} {\bibinfo {author} {\bibfnamefont {K.}~\bibnamefont
  {Riedl}}, \bibinfo {author} {\bibfnamefont {E.}~\bibnamefont {Gati}},\ and\
  \bibinfo {author} {\bibfnamefont {R.}~\bibnamefont {Valent{'i}}},\ }\bibfield
   {title} {\bibinfo {title} {Ingredients for generalized models of
  $\kappa$-phase organic charge-transfer salts: A review},\ }\href
  {https://doi.org/10.3390/cryst12121689} {\bibfield  {journal} {\bibinfo
  {journal} {Crystals}\ }\textbf {\bibinfo {volume} {12}},\ \bibinfo {eid}
  {1689} (\bibinfo {year} {2022})}\BibitemShut {NoStop}%
\bibitem [{\citenamefont {Liebman}\ \emph {et~al.}(2024)\citenamefont
  {Liebman}, \citenamefont {Miyagawa}, \citenamefont {Kanoda},\ and\
  \citenamefont {Drichko}}]{Liebman2024}%
  \BibitemOpen
  \bibfield  {author} {\bibinfo {author} {\bibfnamefont {J.}~\bibnamefont
  {Liebman}}, \bibinfo {author} {\bibfnamefont {K.}~\bibnamefont {Miyagawa}},
  \bibinfo {author} {\bibfnamefont {K.}~\bibnamefont {Kanoda}},\ and\ \bibinfo
  {author} {\bibfnamefont {N.}~\bibnamefont {Drichko}},\ }\href
  {https://arxiv.org/abs/2403.02676} {\bibinfo {title} {Novel dipole-lattice
  coupling in the quantum-spin-liquid material
  $\kappa$-{{(BEDT-TTF)$_2$Cu$_2$(CN)$_3$}}}} (\bibinfo {year} {2024}),\
  \Eprint {https://arxiv.org/abs/2403.02676} {arXiv:2403.02676} \BibitemShut
  {NoStop}%
\bibitem [{\citenamefont {Kandpal}\ \emph {et~al.}(2009)\citenamefont
  {Kandpal}, \citenamefont {Opahle}, \citenamefont {Zhang}, \citenamefont
  {Jeschke},\ and\ \citenamefont {Valent{\'{i}}}}]{Kandpal2009}%
  \BibitemOpen
  \bibfield  {author} {\bibinfo {author} {\bibfnamefont {H.~C.}\ \bibnamefont
  {Kandpal}}, \bibinfo {author} {\bibfnamefont {I.}~\bibnamefont {Opahle}},
  \bibinfo {author} {\bibfnamefont {Y.-Z.}\ \bibnamefont {Zhang}}, \bibinfo
  {author} {\bibfnamefont {H.~O.}\ \bibnamefont {Jeschke}},\ and\ \bibinfo
  {author} {\bibfnamefont {R.}~\bibnamefont {Valent{\'{i}}}},\ }\bibfield
  {title} {\bibinfo {title} {Revision of model parameters for $\kappa$-type
  charge transfer salts: An ab initio study},\ }\href
  {https://link.aps.org/doi/10.1103/PhysRevLett.103.067004} {\bibfield
  {journal} {\bibinfo  {journal} {Phys. Rev. Lett.}\ }\textbf {\bibinfo
  {volume} {103}},\ \bibinfo {pages} {67004} (\bibinfo {year}
  {2009})}\BibitemShut {NoStop}%
\bibitem [{\citenamefont {Foury-Leylekian}\ \emph {et~al.}(2020)\citenamefont
  {Foury-Leylekian}, \citenamefont {Ilakovac}, \citenamefont {Fertey},
  \citenamefont {Baledent}, \citenamefont {Milat}, \citenamefont {Miyagawa},
  \citenamefont {Kanoda}, \citenamefont {Hiramatsu}, \citenamefont {Yoshida},
  \citenamefont {Saito}, \citenamefont {Alemany}, \citenamefont {Canadell},
  \citenamefont {Tomic},\ and\ \citenamefont {Pouget}}]{Leylekian2020}%
  \BibitemOpen
  \bibfield  {author} {\bibinfo {author} {\bibfnamefont {P.}~\bibnamefont
  {Foury-Leylekian}}, \bibinfo {author} {\bibfnamefont {V.}~\bibnamefont
  {Ilakovac}}, \bibinfo {author} {\bibfnamefont {P.}~\bibnamefont {Fertey}},
  \bibinfo {author} {\bibfnamefont {V.}~\bibnamefont {Baledent}}, \bibinfo
  {author} {\bibfnamefont {O.}~\bibnamefont {Milat}}, \bibinfo {author}
  {\bibfnamefont {K.}~\bibnamefont {Miyagawa}}, \bibinfo {author}
  {\bibfnamefont {K.}~\bibnamefont {Kanoda}}, \bibinfo {author} {\bibfnamefont
  {T.}~\bibnamefont {Hiramatsu}}, \bibinfo {author} {\bibfnamefont
  {Y.}~\bibnamefont {Yoshida}}, \bibinfo {author} {\bibfnamefont
  {G.}~\bibnamefont {Saito}}, \bibinfo {author} {\bibfnamefont
  {P.}~\bibnamefont {Alemany}}, \bibinfo {author} {\bibfnamefont
  {E.}~\bibnamefont {Canadell}}, \bibinfo {author} {\bibfnamefont
  {S.}~\bibnamefont {Tomic}},\ and\ \bibinfo {author} {\bibfnamefont {J.-P.}\
  \bibnamefont {Pouget}},\ }\bibfield  {title} {\bibinfo {title} {{New insights
  into the structural properties of $\kappa$-{{(BEDT-TTF)$_2$Ag$_2$(CN)$_3$}}
  spin liquid}},\ }\href {https://doi.org/10.1107/S2052520620005545} {\bibfield
   {journal} {\bibinfo  {journal} {Acta Cryst. B}\ }\textbf {\bibinfo {volume}
  {76}},\ \bibinfo {pages} {581} (\bibinfo {year} {2020})}\BibitemShut
  {NoStop}%
\bibitem [{\citenamefont {Hartmann}\ \emph {et~al.}(2019)\citenamefont
  {Hartmann}, \citenamefont {Gati}, \citenamefont {Yoshida}, \citenamefont
  {Saito},\ and\ \citenamefont {Lang}}]{Hartmann2019}%
  \BibitemOpen
  \bibfield  {author} {\bibinfo {author} {\bibfnamefont {S.}~\bibnamefont
  {Hartmann}}, \bibinfo {author} {\bibfnamefont {E.}~\bibnamefont {Gati}},
  \bibinfo {author} {\bibfnamefont {Y.}~\bibnamefont {Yoshida}}, \bibinfo
  {author} {\bibfnamefont {G.}~\bibnamefont {Saito}},\ and\ \bibinfo {author}
  {\bibfnamefont {M.}~\bibnamefont {Lang}},\ }\bibfield  {title} {\bibinfo
  {title} {Thermal expansion studies on the spin-liquid-candidate system
  $\kappa$-{{(BEDT-TTF)$_2$Ag$_2$(CN)$_3$}}},\ }\href
  {https://doi.org/10.1002/pssb.201800640} {\bibfield  {journal} {\bibinfo
  {journal} {phys. stat. sol. (b)}\ }\textbf {\bibinfo {volume} {256}},\
  \bibinfo {pages} {1800640} (\bibinfo {year} {2019})}\BibitemShut {NoStop}%
\bibitem [{\citenamefont {Hiramatsu}\ \emph {et~al.}(2017)\citenamefont
  {Hiramatsu}, \citenamefont {Yoshida}, \citenamefont {Saito}, \citenamefont
  {Otsuka}, \citenamefont {Yamochi}, \citenamefont {Maesato}, \citenamefont
  {Shimizu}, \citenamefont {Ito}, \citenamefont {Nakamura}, \citenamefont
  {Kishida}, \citenamefont {Watanabe},\ and\ \citenamefont
  {Kumai}}]{Hiramatsu2017}%
  \BibitemOpen
  \bibfield  {author} {\bibinfo {author} {\bibfnamefont {T.}~\bibnamefont
  {Hiramatsu}}, \bibinfo {author} {\bibfnamefont {Y.}~\bibnamefont {Yoshida}},
  \bibinfo {author} {\bibfnamefont {G.}~\bibnamefont {Saito}}, \bibinfo
  {author} {\bibfnamefont {A.}~\bibnamefont {Otsuka}}, \bibinfo {author}
  {\bibfnamefont {H.}~\bibnamefont {Yamochi}}, \bibinfo {author} {\bibfnamefont
  {M.}~\bibnamefont {Maesato}}, \bibinfo {author} {\bibfnamefont
  {Y.}~\bibnamefont {Shimizu}}, \bibinfo {author} {\bibfnamefont
  {H.}~\bibnamefont {Ito}}, \bibinfo {author} {\bibfnamefont {Y.}~\bibnamefont
  {Nakamura}}, \bibinfo {author} {\bibfnamefont {H.}~\bibnamefont {Kishida}},
  \bibinfo {author} {\bibfnamefont {M.}~\bibnamefont {Watanabe}},\ and\
  \bibinfo {author} {\bibfnamefont {R.}~\bibnamefont {Kumai}},\ }\bibfield
  {title} {\bibinfo {title} {Design and preparation of a quantum spin liquid
  candidate {{$\kappa$-(ET)$_2$Ag$_2$(CN)$_3$}} having a nearby
  superconductivity},\ }\href {https://doi.org/10.1246/bcsj.20170167}
  {\bibfield  {journal} {\bibinfo  {journal} {Bull. Chem. Soc. Jpn.}\ }\textbf
  {\bibinfo {volume} {90}},\ \bibinfo {pages} {1073} (\bibinfo {year}
  {2017})}\BibitemShut {NoStop}%
\bibitem [{\citenamefont {R\"osslhuber}\ \emph {et~al.}(2023)\citenamefont
  {R\"osslhuber}, \citenamefont {H\"ubner}, \citenamefont {Dressel},\ and\
  \citenamefont {Pustogow}}]{Rosslhuber2023}%
  \BibitemOpen
  \bibfield  {author} {\bibinfo {author} {\bibfnamefont {R.}~\bibnamefont
  {R\"osslhuber}}, \bibinfo {author} {\bibfnamefont {R.}~\bibnamefont
  {H\"ubner}}, \bibinfo {author} {\bibfnamefont {M.}~\bibnamefont {Dressel}},\
  and\ \bibinfo {author} {\bibfnamefont {A.}~\bibnamefont {Pustogow}},\
  }\bibfield  {title} {\bibinfo {title} {Pressure-dependent dielectric response
  of the frustrated {{Mott}} insulator
  $\kappa$-{{(BEDT-TTF)$_{2}$Ag$_{2}$(CN)$_{3}$}}},\ }\href
  {https://doi.org/10.1103/PhysRevB.107.075113} {\bibfield  {journal} {\bibinfo
   {journal} {Phys. Rev. B}\ }\textbf {\bibinfo {volume} {107}},\ \bibinfo
  {pages} {075113} (\bibinfo {year} {2023})}\BibitemShut {NoStop}%
\bibitem [{\citenamefont {Els{\"{a}}sser}\ \emph {et~al.}(2012)\citenamefont
  {Els{\"{a}}sser}, \citenamefont {Wu}, \citenamefont {Dressel},\ and\
  \citenamefont {Schlueter}}]{Elsasser2012}%
  \BibitemOpen
  \bibfield  {author} {\bibinfo {author} {\bibfnamefont {S.}~\bibnamefont
  {Els{\"{a}}sser}}, \bibinfo {author} {\bibfnamefont {D.}~\bibnamefont {Wu}},
  \bibinfo {author} {\bibfnamefont {M.}~\bibnamefont {Dressel}},\ and\ \bibinfo
  {author} {\bibfnamefont {J.~A.}\ \bibnamefont {Schlueter}},\ }\bibfield
  {title} {\bibinfo {title} {{Power-law dependence of the optical conductivity
  observed in the quantum spin-liquid compound
  $\ensuremath{\kappa}$-(BEDT-TTF)${}_{2}$Cu${}_{2}$(CN)${}_{3}$}},\ }\href
  {https://link.aps.org/doi/10.1103/PhysRevB.86.155150} {\bibfield  {journal}
  {\bibinfo  {journal} {Phys. Rev. B}\ }\textbf {\bibinfo {volume} {86}},\
  \bibinfo {pages} {155150} (\bibinfo {year} {2012})}\BibitemShut {NoStop}%
\bibitem [{SM()}]{SM}%
  \BibitemOpen
  \href@noop {} {}\bibinfo {note} {Details on the measurement technique are
  given in the Supplemental Materials. In addition to the raw date, the sample
  to sample variation is discussed, as well as the angular
  dependence.}\BibitemShut {Stop}%
\bibitem [{\citenamefont {Pinteri{\'{c}}}\ \emph {et~al.}(2016)\citenamefont
  {Pinteri{\'{c}}}, \citenamefont {Lazi{\'{c}}}, \citenamefont {Pustogow},
  \citenamefont {Ivek}, \citenamefont {Kuve{\v{z}}di{\'{c}}}, \citenamefont
  {Milat}, \citenamefont {Gumhalter}, \citenamefont {Basleti{\'{c}}},
  \citenamefont {{\v{C}}ulo}, \citenamefont {Korin-Hamzi{\'{c}}}, \citenamefont
  {L{\"{o}}hle}, \citenamefont {H{\"{u}}bner}, \citenamefont {{Sanz Alonso}},
  \citenamefont {Hiramatsu}, \citenamefont {Yoshida}, \citenamefont {Saito},
  \citenamefont {Dressel},\ and\ \citenamefont {Tomi{\'{c}}}}]{Pinteric2016}%
  \BibitemOpen
  \bibfield  {author} {\bibinfo {author} {\bibfnamefont {M.}~\bibnamefont
  {Pinteri{\'{c}}}}, \bibinfo {author} {\bibfnamefont {P.}~\bibnamefont
  {Lazi{\'{c}}}}, \bibinfo {author} {\bibfnamefont {A.}~\bibnamefont
  {Pustogow}}, \bibinfo {author} {\bibfnamefont {T.}~\bibnamefont {Ivek}},
  \bibinfo {author} {\bibfnamefont {M.}~\bibnamefont {Kuve{\v{z}}di{\'{c}}}},
  \bibinfo {author} {\bibfnamefont {O.}~\bibnamefont {Milat}}, \bibinfo
  {author} {\bibfnamefont {B.}~\bibnamefont {Gumhalter}}, \bibinfo {author}
  {\bibfnamefont {M.}~\bibnamefont {Basleti{\'{c}}}}, \bibinfo {author}
  {\bibfnamefont {M.}~\bibnamefont {{\v{C}}ulo}}, \bibinfo {author}
  {\bibfnamefont {B.}~\bibnamefont {Korin-Hamzi{\'{c}}}}, \bibinfo {author}
  {\bibfnamefont {A.}~\bibnamefont {L{\"{o}}hle}}, \bibinfo {author}
  {\bibfnamefont {R.}~\bibnamefont {H{\"{u}}bner}}, \bibinfo {author}
  {\bibfnamefont {M.}~\bibnamefont {{Sanz Alonso}}}, \bibinfo {author}
  {\bibfnamefont {T.}~\bibnamefont {Hiramatsu}}, \bibinfo {author}
  {\bibfnamefont {Y.}~\bibnamefont {Yoshida}}, \bibinfo {author} {\bibfnamefont
  {G.}~\bibnamefont {Saito}}, \bibinfo {author} {\bibfnamefont
  {M.}~\bibnamefont {Dressel}},\ and\ \bibinfo {author} {\bibfnamefont
  {S.}~\bibnamefont {Tomi{\'{c}}}},\ }\bibfield  {title} {\bibinfo {title}
  {Anion effects on electronic structure and electrodynamic properties of the
  {{Mott}} insulator $\kappa$-{{(BEDT-TTF)$_{2}$Ag$_{2}$(CN)$_{3}$}}},\ }\href
  {https://link.aps.org/doi/10.1103/PhysRevB.94.161105} {\bibfield  {journal}
  {\bibinfo  {journal} {Phys. Rev. B}\ }\textbf {\bibinfo {volume} {94}},\
  \bibinfo {pages} {161105} (\bibinfo {year} {2016})}\BibitemShut {NoStop}%
\bibitem [{\citenamefont {Elstner}\ \emph {et~al.}(1993)\citenamefont
  {Elstner}, \citenamefont {Singh},\ and\ \citenamefont {Young}}]{Elstner1993}%
  \BibitemOpen
  \bibfield  {author} {\bibinfo {author} {\bibfnamefont {N.}~\bibnamefont
  {Elstner}}, \bibinfo {author} {\bibfnamefont {R.~R.~P.}\ \bibnamefont
  {Singh}},\ and\ \bibinfo {author} {\bibfnamefont {A.~P.}\ \bibnamefont
  {Young}},\ }\bibfield  {title} {\bibinfo {title} {Finite temperature
  properties of the spin-1/2 {{Heisenberg }} antiferromagnet on the triangular
  lattice},\ }\href {https://doi.org/10.1103/PhysRevLett.71.1629} {\bibfield
  {journal} {\bibinfo  {journal} {Phys. Rev. Lett.}\ }\textbf {\bibinfo
  {volume} {71}},\ \bibinfo {pages} {1629} (\bibinfo {year}
  {1993})}\BibitemShut {NoStop}%
\bibitem [{\citenamefont {Pohle}\ and\ \citenamefont
  {Jaubert}(2023)}]{Pohle2023}%
  \BibitemOpen
  \bibfield  {author} {\bibinfo {author} {\bibfnamefont {R.}~\bibnamefont
  {Pohle}}\ and\ \bibinfo {author} {\bibfnamefont {L.~D.~C.}\ \bibnamefont
  {Jaubert}},\ }\bibfield  {title} {\bibinfo {title} {Curie-law crossover in
  spin liquids},\ }\href {https://doi.org/10.1103/PhysRevB.108.024411}
  {\bibfield  {journal} {\bibinfo  {journal} {Phys. Rev. B}\ }\textbf {\bibinfo
  {volume} {108}},\ \bibinfo {pages} {024411} (\bibinfo {year}
  {2023})}\BibitemShut {NoStop}%
\bibitem [{\citenamefont {Furukawa}\ \emph {et~al.}(2015)\citenamefont
  {Furukawa}, \citenamefont {Miyagawa}, \citenamefont {Itou}, \citenamefont
  {Ito}, \citenamefont {Taniguchi}, \citenamefont {Saito}, \citenamefont
  {Iguchi}, \citenamefont {Sasaki},\ and\ \citenamefont
  {Kanoda}}]{Furukawa2015}%
  \BibitemOpen
  \bibfield  {author} {\bibinfo {author} {\bibfnamefont {T.}~\bibnamefont
  {Furukawa}}, \bibinfo {author} {\bibfnamefont {K.}~\bibnamefont {Miyagawa}},
  \bibinfo {author} {\bibfnamefont {T.}~\bibnamefont {Itou}}, \bibinfo {author}
  {\bibfnamefont {M.}~\bibnamefont {Ito}}, \bibinfo {author} {\bibfnamefont
  {H.}~\bibnamefont {Taniguchi}}, \bibinfo {author} {\bibfnamefont
  {M.}~\bibnamefont {Saito}}, \bibinfo {author} {\bibfnamefont
  {S.}~\bibnamefont {Iguchi}}, \bibinfo {author} {\bibfnamefont
  {T.}~\bibnamefont {Sasaki}},\ and\ \bibinfo {author} {\bibfnamefont
  {K.}~\bibnamefont {Kanoda}},\ }\bibfield  {title} {\bibinfo {title} {Quantum
  spin liquid emerging from antiferromagnetic order by introducing disorder},\
  }\href {https://doi.org/10.1103/PhysRevLett.115.077001} {\bibfield  {journal}
  {\bibinfo  {journal} {Phys. Rev. Lett.}\ }\textbf {\bibinfo {volume} {115}},\
  \bibinfo {pages} {077001} (\bibinfo {year} {2015})}\BibitemShut {NoStop}%
\bibitem [{\citenamefont {Kimchi}\ \emph {et~al.}(2018)\citenamefont {Kimchi},
  \citenamefont {Nahum},\ and\ \citenamefont {Senthil}}]{Kimchi2018}%
  \BibitemOpen
  \bibfield  {author} {\bibinfo {author} {\bibfnamefont {I.}~\bibnamefont
  {Kimchi}}, \bibinfo {author} {\bibfnamefont {A.}~\bibnamefont {Nahum}},\ and\
  \bibinfo {author} {\bibfnamefont {T.}~\bibnamefont {Senthil}},\ }\bibfield
  {title} {\bibinfo {title} {Valence bonds in random quantum magnets: Theory
  and application to {{YbMgGaO$_{4}$}}},\ }\href
  {https://doi.org/10.1103/PhysRevX.8.031028} {\bibfield  {journal} {\bibinfo
  {journal} {Phys. Rev. X}\ }\textbf {\bibinfo {volume} {8}},\ \bibinfo {pages}
  {031028} (\bibinfo {year} {2018})}\BibitemShut {NoStop}%
\bibitem [{\citenamefont {Pal}\ \emph {et~al.}(2019)\citenamefont {Pal},
  \citenamefont {Kumar}, \citenamefont {Sharma}, \citenamefont {Banerjee},
  \citenamefont {Roy}, \citenamefont {Park}, \citenamefont {Nigam},\ and\
  \citenamefont {Cheong}}]{Pal2020}%
  \BibitemOpen
  \bibfield  {author} {\bibinfo {author} {\bibfnamefont {S.}~\bibnamefont
  {Pal}}, \bibinfo {author} {\bibfnamefont {K.}~\bibnamefont {Kumar}}, \bibinfo
  {author} {\bibfnamefont {R.}~\bibnamefont {Sharma}}, \bibinfo {author}
  {\bibfnamefont {A.}~\bibnamefont {Banerjee}}, \bibinfo {author}
  {\bibfnamefont {S.~B.}\ \bibnamefont {Roy}}, \bibinfo {author} {\bibfnamefont
  {J.-G.}\ \bibnamefont {Park}}, \bibinfo {author} {\bibfnamefont {A.~K.}\
  \bibnamefont {Nigam}},\ and\ \bibinfo {author} {\bibfnamefont {S.-W.}\
  \bibnamefont {Cheong}},\ }\bibfield  {title} {\bibinfo {title} {Possible
  glass-like random singlet magnetic state in {{1T-TaS$_2$}}},\ }\href
  {https://doi.org/10.1088/1361-648X/ab48be} {\bibfield  {journal} {\bibinfo
  {journal} {J. of Phys.: Condens. Matter}\ }\textbf {\bibinfo {volume} {32}},\
  \bibinfo {pages} {035601} (\bibinfo {year} {2019})}\BibitemShut {NoStop}%
\bibitem [{\citenamefont {Pal}\ and\ \citenamefont {Roy}(2022)}]{Pal2022}%
  \BibitemOpen
  \bibfield  {author} {\bibinfo {author} {\bibfnamefont {S.}~\bibnamefont
  {Pal}}\ and\ \bibinfo {author} {\bibfnamefont {S.}~\bibnamefont {Roy}},\
  }\bibfield  {title} {\bibinfo {title} {Understanding the magnetic response of
  the quantum spin liquid compound {{1T-TaS$_2$}}},\ }\href
  {https://doi.org/https://doi.org/10.1016/j.physb.2022.414121} {\bibfield
  {journal} {\bibinfo  {journal} {Physica B}\ }\textbf {\bibinfo {volume}
  {643}},\ \bibinfo {pages} {414121} (\bibinfo {year} {2022})}\BibitemShut
  {NoStop}%
\bibitem [{\citenamefont {Wang}\ \emph {et~al.}(2021)\citenamefont {Wang},
  \citenamefont {Yuan}, \citenamefont {Singer}, \citenamefont {Smaha},
  \citenamefont {He}, \citenamefont {Wen}, \citenamefont {Lee},\ and\
  \citenamefont {Imai}}]{Wang2021}%
  \BibitemOpen
  \bibfield  {author} {\bibinfo {author} {\bibfnamefont {J.}~\bibnamefont
  {Wang}}, \bibinfo {author} {\bibfnamefont {W.}~\bibnamefont {Yuan}}, \bibinfo
  {author} {\bibfnamefont {P.~M.}\ \bibnamefont {Singer}}, \bibinfo {author}
  {\bibfnamefont {R.~W.}\ \bibnamefont {Smaha}}, \bibinfo {author}
  {\bibfnamefont {W.}~\bibnamefont {He}}, \bibinfo {author} {\bibfnamefont
  {J.}~\bibnamefont {Wen}}, \bibinfo {author} {\bibfnamefont {Y.~S.}\
  \bibnamefont {Lee}},\ and\ \bibinfo {author} {\bibfnamefont {T.}~\bibnamefont
  {Imai}},\ }\bibfield  {title} {\bibinfo {title} {Emergence of spin singlets
  with inhomogeneous gaps in the kagome lattice {{Heisenberg }}
  antiferromagnets {{Zn}}-barlowite and herbertsmithite},\ }\href
  {https://doi.org/10.1038/s41567-021-01310-3} {\bibfield  {journal} {\bibinfo
  {journal} {Nat. Phys.}\ }\textbf {\bibinfo {volume} {17}},\ \bibinfo {pages}
  {1109} (\bibinfo {year} {2021})}\BibitemShut {NoStop}%
\bibitem [{\citenamefont {Chabre}\ \emph {et~al.}(2005)\citenamefont {Chabre},
  \citenamefont {Ghorayeb}, \citenamefont {Millet}, \citenamefont
  {Pashchenko},\ and\ \citenamefont {Stepanov}}]{Chabre2005}%
  \BibitemOpen
  \bibfield  {author} {\bibinfo {author} {\bibfnamefont {F.}~\bibnamefont
  {Chabre}}, \bibinfo {author} {\bibfnamefont {A.~M.}\ \bibnamefont
  {Ghorayeb}}, \bibinfo {author} {\bibfnamefont {P.}~\bibnamefont {Millet}},
  \bibinfo {author} {\bibfnamefont {V.~A.}\ \bibnamefont {Pashchenko}},\ and\
  \bibinfo {author} {\bibfnamefont {A.}~\bibnamefont {Stepanov}},\ }\bibfield
  {title} {\bibinfo {title} {Low-temperature behavior of the {{ESR }} linewidth
  in a system with a spin gap: $\eta$-{{Na$_{1.286}$V$_{2}$O$_{5}$}}},\ }\href
  {https://doi.org/10.1103/PhysRevB.72.012415} {\bibfield  {journal} {\bibinfo
  {journal} {Phys. Rev. B}\ }\textbf {\bibinfo {volume} {72}},\ \bibinfo
  {pages} {012415} (\bibinfo {year} {2005})}\BibitemShut {NoStop}%
\bibitem [{\citenamefont {Deisenhofer}\ \emph {et~al.}(2006)\citenamefont
  {Deisenhofer}, \citenamefont {Eremina}, \citenamefont {Pimenov},
  \citenamefont {Gavrilova}, \citenamefont {Berger}, \citenamefont {Johnsson},
  \citenamefont {Lemmens}, \citenamefont {Krug~von Nidda}, \citenamefont
  {Loidl}, \citenamefont {Lee},\ and\ \citenamefont
  {Whangbo}}]{Deisenhofer2006}%
  \BibitemOpen
  \bibfield  {author} {\bibinfo {author} {\bibfnamefont {J.}~\bibnamefont
  {Deisenhofer}}, \bibinfo {author} {\bibfnamefont {R.~M.}\ \bibnamefont
  {Eremina}}, \bibinfo {author} {\bibfnamefont {A.}~\bibnamefont {Pimenov}},
  \bibinfo {author} {\bibfnamefont {T.}~\bibnamefont {Gavrilova}}, \bibinfo
  {author} {\bibfnamefont {H.}~\bibnamefont {Berger}}, \bibinfo {author}
  {\bibfnamefont {M.}~\bibnamefont {Johnsson}}, \bibinfo {author}
  {\bibfnamefont {P.}~\bibnamefont {Lemmens}}, \bibinfo {author} {\bibfnamefont
  {H.-A.}\ \bibnamefont {Krug~von Nidda}}, \bibinfo {author} {\bibfnamefont
  {A.}~\bibnamefont {Loidl}}, \bibinfo {author} {\bibfnamefont {K.-S.}\
  \bibnamefont {Lee}},\ and\ \bibinfo {author} {\bibfnamefont {M.-H.}\
  \bibnamefont {Whangbo}},\ }\bibfield  {title} {\bibinfo {title} {Structural
  and magnetic dimers in the spin-gapped system {{CuTe$_{2}$O$_{5}$}}},\ }\href
  {https://doi.org/10.1103/PhysRevB.74.174421} {\bibfield  {journal} {\bibinfo
  {journal} {Phys. Rev. B}\ }\textbf {\bibinfo {volume} {74}},\ \bibinfo
  {pages} {174421} (\bibinfo {year} {2006})}\BibitemShut {NoStop}%
\bibitem [{\citenamefont {Vasiliev}\ \emph {et~al.}(2015)\citenamefont
  {Vasiliev}, \citenamefont {Volkova}, \citenamefont {Zvereva}, \citenamefont
  {Koshelev}, \citenamefont {Urusov}, \citenamefont {Chareev}, \citenamefont
  {Petkov}, \citenamefont {Sukhanov}, \citenamefont {Rahaman},\ and\
  \citenamefont {Saha-Dasgupta}}]{Vasiliev2015}%
  \BibitemOpen
  \bibfield  {author} {\bibinfo {author} {\bibfnamefont {A.~N.}\ \bibnamefont
  {Vasiliev}}, \bibinfo {author} {\bibfnamefont {O.~S.}\ \bibnamefont
  {Volkova}}, \bibinfo {author} {\bibfnamefont {E.~A.}\ \bibnamefont
  {Zvereva}}, \bibinfo {author} {\bibfnamefont {A.~V.}\ \bibnamefont
  {Koshelev}}, \bibinfo {author} {\bibfnamefont {V.~S.}\ \bibnamefont
  {Urusov}}, \bibinfo {author} {\bibfnamefont {D.~A.}\ \bibnamefont {Chareev}},
  \bibinfo {author} {\bibfnamefont {V.~I.}\ \bibnamefont {Petkov}}, \bibinfo
  {author} {\bibfnamefont {M.~V.}\ \bibnamefont {Sukhanov}}, \bibinfo {author}
  {\bibfnamefont {B.}~\bibnamefont {Rahaman}},\ and\ \bibinfo {author}
  {\bibfnamefont {T.}~\bibnamefont {Saha-Dasgupta}},\ }\bibfield  {title}
  {\bibinfo {title} {Valence-bond solid as the quantum ground state in
  honeycomb layered urusovite {{CuAl(AsO$_{4}$)O}}},\ }\href
  {https://doi.org/10.1103/PhysRevB.91.144406} {\bibfield  {journal} {\bibinfo
  {journal} {Phys. Rev. B}\ }\textbf {\bibinfo {volume} {91}},\ \bibinfo
  {pages} {144406} (\bibinfo {year} {2015})}\BibitemShut {NoStop}%
\bibitem [{\citenamefont {Pustogow}\ \emph {et~al.}(2020)\citenamefont
  {Pustogow}, \citenamefont {Le}, \citenamefont {Wang}, \citenamefont {Luo},
  \citenamefont {Gati}, \citenamefont {Schubert}, \citenamefont {Lang},\ and\
  \citenamefont {Brown}}]{Pustogow2020}%
  \BibitemOpen
  \bibfield  {author} {\bibinfo {author} {\bibfnamefont {A.}~\bibnamefont
  {Pustogow}}, \bibinfo {author} {\bibfnamefont {T.}~\bibnamefont {Le}},
  \bibinfo {author} {\bibfnamefont {H.-H.}\ \bibnamefont {Wang}}, \bibinfo
  {author} {\bibfnamefont {Y.}~\bibnamefont {Luo}}, \bibinfo {author}
  {\bibfnamefont {E.}~\bibnamefont {Gati}}, \bibinfo {author} {\bibfnamefont
  {H.}~\bibnamefont {Schubert}}, \bibinfo {author} {\bibfnamefont
  {M.}~\bibnamefont {Lang}},\ and\ \bibinfo {author} {\bibfnamefont {S.~E.}\
  \bibnamefont {Brown}},\ }\bibfield  {title} {\bibinfo {title} {{Impurity
  moments conceal low-energy relaxation of quantum spin liquids}},\ }\href
  {https://doi.org/10.1103/PhysRevB.101.140401} {\bibfield  {journal} {\bibinfo
   {journal} {Phys. Rev. B}\ }\textbf {\bibinfo {volume} {101}},\ \bibinfo
  {pages} {140401(R)} (\bibinfo {year} {2020})}\BibitemShut {NoStop}%
\bibitem [{\citenamefont {Zorko}\ \emph {et~al.}(2017)\citenamefont {Zorko},
  \citenamefont {Herak}, \citenamefont {Gomil\ifmmode~\check{s}\else
  \v{s}\fi{}ek}, \citenamefont {van Tol}, \citenamefont {Vel\'azquez},
  \citenamefont {Khuntia}, \citenamefont {Bert},\ and\ \citenamefont
  {Mendels}}]{Zorko2017}%
  \BibitemOpen
  \bibfield  {author} {\bibinfo {author} {\bibfnamefont {A.}~\bibnamefont
  {Zorko}}, \bibinfo {author} {\bibfnamefont {M.}~\bibnamefont {Herak}},
  \bibinfo {author} {\bibfnamefont {M.}~\bibnamefont
  {Gomil\ifmmode~\check{s}\else \v{s}\fi{}ek}}, \bibinfo {author}
  {\bibfnamefont {J.}~\bibnamefont {van Tol}}, \bibinfo {author} {\bibfnamefont
  {M.}~\bibnamefont {Vel\'azquez}}, \bibinfo {author} {\bibfnamefont
  {P.}~\bibnamefont {Khuntia}}, \bibinfo {author} {\bibfnamefont
  {F.}~\bibnamefont {Bert}},\ and\ \bibinfo {author} {\bibfnamefont
  {P.}~\bibnamefont {Mendels}},\ }\bibfield  {title} {\bibinfo {title}
  {Symmetry reduction in the quantum kagome antiferromagnet herbertsmithite},\
  }\href {https://doi.org/10.1103/PhysRevLett.118.017202} {\bibfield  {journal}
  {\bibinfo  {journal} {Phys. Rev. Lett.}\ }\textbf {\bibinfo {volume} {118}},\
  \bibinfo {pages} {017202} (\bibinfo {year} {2017})}\BibitemShut {NoStop}%
\bibitem [{\citenamefont {Yang}\ \emph {et~al.}(2023)\citenamefont {Yang},
  \citenamefont {Goh}, \citenamefont {Sung}, \citenamefont {Ye}, \citenamefont
  {Biswas}, \citenamefont {Kaib}, \citenamefont {Dhakal}, \citenamefont {Yan},
  \citenamefont {Li}, \citenamefont {Jiang}, \citenamefont {Chen},
  \citenamefont {Lei}, \citenamefont {He}, \citenamefont {Valent\'{i}},
  \citenamefont {Winter}, \citenamefont {Hovden},\ and\ \citenamefont
  {Tsen}}]{Yang2023}%
  \BibitemOpen
  \bibfield  {author} {\bibinfo {author} {\bibfnamefont {B.}~\bibnamefont
  {Yang}}, \bibinfo {author} {\bibfnamefont {Y.~M.}\ \bibnamefont {Goh}},
  \bibinfo {author} {\bibfnamefont {S.~H.}\ \bibnamefont {Sung}}, \bibinfo
  {author} {\bibfnamefont {G.}~\bibnamefont {Ye}}, \bibinfo {author}
  {\bibfnamefont {S.}~\bibnamefont {Biswas}}, \bibinfo {author} {\bibfnamefont
  {D.~A.~S.}\ \bibnamefont {Kaib}}, \bibinfo {author} {\bibfnamefont
  {R.}~\bibnamefont {Dhakal}}, \bibinfo {author} {\bibfnamefont
  {S.}~\bibnamefont {Yan}}, \bibinfo {author} {\bibfnamefont {C.}~\bibnamefont
  {Li}}, \bibinfo {author} {\bibfnamefont {S.}~\bibnamefont {Jiang}}, \bibinfo
  {author} {\bibfnamefont {F.}~\bibnamefont {Chen}}, \bibinfo {author}
  {\bibfnamefont {H.}~\bibnamefont {Lei}}, \bibinfo {author} {\bibfnamefont
  {R.}~\bibnamefont {He}}, \bibinfo {author} {\bibfnamefont {R.}~\bibnamefont
  {Valent\'{i}}}, \bibinfo {author} {\bibfnamefont {S.~M.}\ \bibnamefont
  {Winter}}, \bibinfo {author} {\bibfnamefont {R.}~\bibnamefont {Hovden}},\
  and\ \bibinfo {author} {\bibfnamefont {A.~W.}\ \bibnamefont {Tsen}},\
  }\bibfield  {title} {\bibinfo {title} {Magnetic anisotropy reversal driven by
  structural symmetry-breaking in monolayer $\alpha$-{{RuCl$_3$}}},\ }\href
  {https://doi.org/10.1038/s41563-022-01401-3} {\bibfield  {journal} {\bibinfo
  {journal} {Nat. Mater.}\ }\textbf {\bibinfo {volume} {22}},\ \bibinfo {pages}
  {50–57} (\bibinfo {year} {2023})}\BibitemShut {NoStop}%
\bibitem [{\citenamefont {Jiang}\ and\ \citenamefont
  {Jiang}(2023)}]{Jiang2023}%
  \BibitemOpen
  \bibfield  {author} {\bibinfo {author} {\bibfnamefont {Y.-F.}\ \bibnamefont
  {Jiang}}\ and\ \bibinfo {author} {\bibfnamefont {H.-C.}\ \bibnamefont
  {Jiang}},\ }\bibfield  {title} {\bibinfo {title} {Nature of quantum spin
  liquids of the $s=\frac{1}{2}$ {{Heisenberg}} antiferromagnet on the
  triangular lattice: A parallel {{DMRG}} study},\ }\href
  {https://doi.org/10.1103/PhysRevB.107.L140411} {\bibfield  {journal}
  {\bibinfo  {journal} {Phys. Rev. B}\ }\textbf {\bibinfo {volume} {107}},\
  \bibinfo {pages} {L140411} (\bibinfo {year} {2023})}\BibitemShut {NoStop}%
\bibitem [{\citenamefont {Sherman}\ \emph {et~al.}(2023)\citenamefont
  {Sherman}, \citenamefont {Dupont},\ and\ \citenamefont
  {Moore}}]{Sherman2023}%
  \BibitemOpen
  \bibfield  {author} {\bibinfo {author} {\bibfnamefont {N.~E.}\ \bibnamefont
  {Sherman}}, \bibinfo {author} {\bibfnamefont {M.}~\bibnamefont {Dupont}},\
  and\ \bibinfo {author} {\bibfnamefont {J.~E.}\ \bibnamefont {Moore}},\
  }\bibfield  {title} {\bibinfo {title} {Spectral function of the
  ${J}_{1}-{J}_{2}$ {{Heisenberg }} model on the triangular lattice},\ }\href
  {https://doi.org/10.1103/PhysRevB.107.165146} {\bibfield  {journal} {\bibinfo
   {journal} {Phys. Rev. B}\ }\textbf {\bibinfo {volume} {107}},\ \bibinfo
  {pages} {165146} (\bibinfo {year} {2023})}\BibitemShut {NoStop}%
\bibitem [{\citenamefont {Szasz}\ \emph {et~al.}(2020)\citenamefont {Szasz},
  \citenamefont {Motruk}, \citenamefont {Zaletel},\ and\ \citenamefont
  {Moore}}]{Szasz2020}%
  \BibitemOpen
  \bibfield  {author} {\bibinfo {author} {\bibfnamefont {A.}~\bibnamefont
  {Szasz}}, \bibinfo {author} {\bibfnamefont {J.}~\bibnamefont {Motruk}},
  \bibinfo {author} {\bibfnamefont {M.~P.}\ \bibnamefont {Zaletel}},\ and\
  \bibinfo {author} {\bibfnamefont {J.~E.}\ \bibnamefont {Moore}},\ }\bibfield
  {title} {\bibinfo {title} {Chiral spin liquid phase of the triangular lattice
  {{Hubbard }} model: A density matrix renormalization group study},\ }\href
  {https://doi.org/10.1103/PhysRevX.10.021042} {\bibfield  {journal} {\bibinfo
  {journal} {Phys. Rev. X}\ }\textbf {\bibinfo {volume} {10}},\ \bibinfo
  {pages} {021042} (\bibinfo {year} {2020})}\BibitemShut {NoStop}%
\bibitem [{\citenamefont {Szasz}\ and\ \citenamefont
  {Motruk}(2021)}]{Szasz2021}%
  \BibitemOpen
  \bibfield  {author} {\bibinfo {author} {\bibfnamefont {A.}~\bibnamefont
  {Szasz}}\ and\ \bibinfo {author} {\bibfnamefont {J.}~\bibnamefont {Motruk}},\
  }\bibfield  {title} {\bibinfo {title} {Phase diagram of the anisotropic
  triangular lattice {{Hubbard }} model},\ }\href
  {https://doi.org/10.1103/PhysRevB.103.235132} {\bibfield  {journal} {\bibinfo
   {journal} {Phys. Rev. B}\ }\textbf {\bibinfo {volume} {103}},\ \bibinfo
  {pages} {235132} (\bibinfo {year} {2021})}\BibitemShut {NoStop}%
\bibitem [{\citenamefont {Iqbal}\ \emph {et~al.}(2016)\citenamefont {Iqbal},
  \citenamefont {Hu}, \citenamefont {Thomale}, \citenamefont {Poilblanc},\ and\
  \citenamefont {Becca}}]{Iqbal2016}%
  \BibitemOpen
  \bibfield  {author} {\bibinfo {author} {\bibfnamefont {Y.}~\bibnamefont
  {Iqbal}}, \bibinfo {author} {\bibfnamefont {W.-J.}\ \bibnamefont {Hu}},
  \bibinfo {author} {\bibfnamefont {R.}~\bibnamefont {Thomale}}, \bibinfo
  {author} {\bibfnamefont {D.}~\bibnamefont {Poilblanc}},\ and\ \bibinfo
  {author} {\bibfnamefont {F.}~\bibnamefont {Becca}},\ }\bibfield  {title}
  {\bibinfo {title} {Spin liquid nature in the {{Heisenberg}}
  ${J}_{1}\ensuremath{-}{J}_{2}$ triangular antiferromagnet},\ }\href
  {https://doi.org/10.1103/PhysRevB.93.144411} {\bibfield  {journal} {\bibinfo
  {journal} {Phys. Rev. B}\ }\textbf {\bibinfo {volume} {93}},\ \bibinfo
  {pages} {144411} (\bibinfo {year} {2016})}\BibitemShut {NoStop}%
\bibitem [{\citenamefont {Shirakawa}\ \emph {et~al.}(2017)\citenamefont
  {Shirakawa}, \citenamefont {Tohyama}, \citenamefont {Kokalj}, \citenamefont
  {Sota},\ and\ \citenamefont {Yunoki}}]{Shirakawa2017}%
  \BibitemOpen
  \bibfield  {author} {\bibinfo {author} {\bibfnamefont {T.}~\bibnamefont
  {Shirakawa}}, \bibinfo {author} {\bibfnamefont {T.}~\bibnamefont {Tohyama}},
  \bibinfo {author} {\bibfnamefont {J.}~\bibnamefont {Kokalj}}, \bibinfo
  {author} {\bibfnamefont {S.}~\bibnamefont {Sota}},\ and\ \bibinfo {author}
  {\bibfnamefont {S.}~\bibnamefont {Yunoki}},\ }\bibfield  {title} {\bibinfo
  {title} {Ground-state phase diagram of the triangular lattice {{Hubbard}}
  model by the density-matrix renormalization group method},\ }\href
  {https://doi.org/10.1103/PhysRevB.96.205130} {\bibfield  {journal} {\bibinfo
  {journal} {Phys. Rev. B}\ }\textbf {\bibinfo {volume} {96}},\ \bibinfo
  {pages} {205130} (\bibinfo {year} {2017})}\BibitemShut {NoStop}%
\bibitem [{\citenamefont {Hu}\ \emph {et~al.}(2019)\citenamefont {Hu},
  \citenamefont {Zhu}, \citenamefont {Eggert},\ and\ \citenamefont
  {He}}]{Hu2019}%
  \BibitemOpen
  \bibfield  {author} {\bibinfo {author} {\bibfnamefont {S.}~\bibnamefont
  {Hu}}, \bibinfo {author} {\bibfnamefont {W.}~\bibnamefont {Zhu}}, \bibinfo
  {author} {\bibfnamefont {S.}~\bibnamefont {Eggert}},\ and\ \bibinfo {author}
  {\bibfnamefont {Y.-C.}\ \bibnamefont {He}},\ }\bibfield  {title} {\bibinfo
  {title} {Dirac spin liquid on the spin-$1/2$ triangular {{Heisenberg }}
  antiferromagnet},\ }\href {https://doi.org/10.1103/PhysRevLett.123.207203}
  {\bibfield  {journal} {\bibinfo  {journal} {Phys. Rev. Lett.}\ }\textbf
  {\bibinfo {volume} {123}},\ \bibinfo {pages} {207203} (\bibinfo {year}
  {2019})}\BibitemShut {NoStop}%
\bibitem [{\citenamefont {Zhu}\ and\ \citenamefont {White}(2015)}]{Zhu2015}%
  \BibitemOpen
  \bibfield  {author} {\bibinfo {author} {\bibfnamefont {Z.}~\bibnamefont
  {Zhu}}\ and\ \bibinfo {author} {\bibfnamefont {S.~R.}\ \bibnamefont
  {White}},\ }\bibfield  {title} {\bibinfo {title} {Spin liquid phase of the
  $s=\frac{1}{2} {J}_{1} \ensuremath{-} {J}_{2}$ {{Heisenberg}} model on the
  triangular lattice},\ }\href {https://doi.org/10.1103/PhysRevB.92.041105}
  {\bibfield  {journal} {\bibinfo  {journal} {Phys. Rev. B}\ }\textbf {\bibinfo
  {volume} {92}},\ \bibinfo {pages} {041105} (\bibinfo {year}
  {2015})}\BibitemShut {NoStop}%
\bibitem [{\citenamefont {Hu}\ \emph {et~al.}(2015)\citenamefont {Hu},
  \citenamefont {Gong}, \citenamefont {Zhu},\ and\ \citenamefont
  {Sheng}}]{Hu2015}%
  \BibitemOpen
  \bibfield  {author} {\bibinfo {author} {\bibfnamefont {W.-J.}\ \bibnamefont
  {Hu}}, \bibinfo {author} {\bibfnamefont {S.-S.}\ \bibnamefont {Gong}},
  \bibinfo {author} {\bibfnamefont {W.}~\bibnamefont {Zhu}},\ and\ \bibinfo
  {author} {\bibfnamefont {D.~N.}\ \bibnamefont {Sheng}},\ }\bibfield  {title}
  {\bibinfo {title} {Competing spin-liquid states in the spin-$\frac{1}{2}$
  {{Heisenberg }} model on the triangular lattice},\ }\href
  {https://doi.org/10.1103/PhysRevB.92.140403} {\bibfield  {journal} {\bibinfo
  {journal} {Phys. Rev. B}\ }\textbf {\bibinfo {volume} {92}},\ \bibinfo
  {pages} {140403} (\bibinfo {year} {2015})}\BibitemShut {NoStop}%
\bibitem [{\citenamefont {Chen}\ \emph {et~al.}(2022)\citenamefont {Chen},
  \citenamefont {Chen}, \citenamefont {Gong}, \citenamefont {Sheng},
  \citenamefont {Li},\ and\ \citenamefont {Weichselbaum}}]{Chen2022}%
  \BibitemOpen
  \bibfield  {author} {\bibinfo {author} {\bibfnamefont {B.-B.}\ \bibnamefont
  {Chen}}, \bibinfo {author} {\bibfnamefont {Z.}~\bibnamefont {Chen}}, \bibinfo
  {author} {\bibfnamefont {S.-S.}\ \bibnamefont {Gong}}, \bibinfo {author}
  {\bibfnamefont {D.~N.}\ \bibnamefont {Sheng}}, \bibinfo {author}
  {\bibfnamefont {W.}~\bibnamefont {Li}},\ and\ \bibinfo {author}
  {\bibfnamefont {A.}~\bibnamefont {Weichselbaum}},\ }\bibfield  {title}
  {\bibinfo {title} {Quantum spin liquid with emergent chiral order in the
  triangular-lattice {{Hubbard }} model},\ }\href
  {https://doi.org/10.1103/PhysRevB.106.094420} {\bibfield  {journal} {\bibinfo
   {journal} {Phys. Rev. B}\ }\textbf {\bibinfo {volume} {106}},\ \bibinfo
  {pages} {094420} (\bibinfo {year} {2022})}\BibitemShut {NoStop}%
\end{thebibliography}
%

\end{document}